\documentclass[%
pra,
 reprint,
 onecolumn,
 aps
]{revtex4-2}
\usepackage{graphicx}
\usepackage{dcolumn}
\usepackage{bm}
\usepackage{float}
\usepackage{amsmath,amssymb,amsfonts}
\usepackage{todonotes}
\usepackage{epstopdf}
\usepackage{diagbox}
\usepackage{multirow}

\def\XXint#1#2#3{{\setbox0=\hbox{$#1{#2#3}{\int}$}
\vcenter{\hbox{$#2#3$}}\kern-.5\wd0}}

\def\Xiint#1{\mathchoice
	{\XXiint\displaystyle\textstyle{#1}}%
	{\XXiint\textstyle\scriptstyle{#1}}%
	{\XXiint\scriptstyle\scriptscriptstyle{#1}}%
	{\XXiint\scriptscriptstyle\scriptscriptstyle{#1}}%
	\!\iint}
\def\XXiint#1#2#3{{\setbox0=\hbox{$#1{#2{\mathrm{#3\!\! #3}}}{\iint}$}
\vcenter{\hbox{$#2 {\mathrm{#3\!\! #3}}$}}\kern-.5\wd0}}
\usepackage{hyperref}
\usepackage{color}  
\definecolor{darkGreen}{rgb}{0,0.45,0}
\definecolor{darkBlue}{rgb}{0,0,0.7}
\definecolor{darkRed}{rgb}{0.76, 0.13, 0.28}
\hypersetup{
	colorlinks=true, 
	linktoc=all,    
	linkcolor=darkBlue, 
	citecolor=darkGreen,
	urlcolor = darkRed,
}
\usepackage{comment}

\renewcommand{\bf}[1]{\mathbf{#1}}
\renewcommand{\d}{{\mathrm{d}}}

\newcommand{\nn}{\noindent}

\usepackage[margin=2.3cm]{geometry}
\usepackage{placeins}
\usepackage[%
linewidth=1pt,
linecolor=red,
topline = false,
rightline = false,
bottomline = false,
rightmargin=0pt,
skipabove=0pt,
skipbelow=0pt,
leftmargin=-.5cm,
innerleftmargin=.5cm,
skipabove=0.3\baselineskip,%
innertopmargin=0.4\baselineskip,%
innerbottommargin=0.4\baselineskip,
innerrightmargin=0pt
]{mdframed}

\begin{document}

\title{Spanwise variations in membrane flutter dynamics}

\author{Christiana Mavroyiakoumou}%
\email[Electronic address: ]{christiana.mav@nyu.edu}
\affiliation{Courant Institute, Applied Math Lab, New York University, New York, NY 10012, USA}

\author{Silas Alben}
\email[Electronic address: ]{alben@umich.edu}
\affiliation{Department of Mathematics, University of Michigan, Ann Arbor, MI 48109, USA}

\date{\today}

\begin{abstract}
We study the large-amplitude flutter of rectangular membranes in 3-D inviscid flows.
The membranes' deformations vary significantly in both the chordwise and spanwise directions. Many previous studies used 2D flow models and neglected spanwise variations, so here we focus on cases with significant spanwise nonuniformity.
We determine when such cases occur and how the dynamics vary over the parameter space of membrane mass and pretension for two sets of boundary conditions and two values of both the Poisson ratio and the membrane aspect ratio.
With spanwise symmetric and asymmetric initial perturbations, the motions differ for long times but eventually reach the same steady state in most cases.

At large times, spanwise symmetric and asymmetric oscillations are seen, with the latter more common. Oscillations are often in the form of ``side-to-side" and other standing wave motions along the span, as well as traveling wave motions, particularly with free side-edges. Motions are generally nonperiodic and more spatially complex with a large membrane mass, and sometimes periodic at small-to-moderate membrane mass.
A large Poisson ratio gives somewhat smoother spatial and temporal features in the dynamics at a given pretension.
Increasing the aspect ratio makes the deflection more uniform along the span.
With different chordwise and spanwise pretensions we find motions that are qualitatively similar to cases with isotropic pretensions between the anisotropic values.
\end{abstract}

\maketitle

\section{Introduction}\label{sec:intro}

The physical mechanisms that govern the interactions between flexible structures and fluid flows can give us key insights into biolocomotion. For instance, scientists have investigated the effects of chordwise and spanwise flexibility of the wings of insects~\cite{wootton1981support,shyy2010recent,eldredge2010roles,reid2019wing}, birds~\cite{yang2012effects,kodali2017effects}, and bats~\cite{swartz1996mechanical,shyy1999flapping,swartz2007wing,song2008aeromechanics,cheney2014membrane,cheney2015wrinkle,yu2020decoupled} and of fins of fish~\cite{lauder2007fishloco} and aquatic mammals~\cite{liu1997propulsive}, which naturally exploit intricate variations of flexibility in both directions to enhance their flapping propulsion performance~\cite{triantafyllou2000hydrodynamics,heathcote2008effect,lin2023performance,smyth2023generating}. Here we focus on very thin and lightweight surfaces that can be modeled as extensible membranes, i.e., soft materials with negligible bending modulus that can significantly stretch in a fluid flow. Examples include rubber, textile fabric, and the skin of swimming and flying animals. 

Unlike a rigid wing that may experience flow separation at its upper surface and significant reductions in aerodynamic efficiency in both steady and unsteady flows, a flexible membrane wing is capable of deforming quickly and adapting to unsteady airflow conditions, thus preventing flow separation and enhancing aircraft maneuverability. Given this superiority in aerodynamic performance, membranes are used in miniature flapping-wing aircraft~\cite{lian2005numerical}, shape-morphing airfoils~\cite{abdulrahim2005flight,lian2005numerical,hu2008flexible,stanford2008fixed,jaworski2012high,piquee2018aerodynamic,schomberg2018transition,tzezana2019thrust}, parachutes~\cite{pepper1971aerodynamic,stein2000parachute}, and sails~\cite{colgate1996fundamentals,kimball2009physics}. An important design consideration for such structures is their stability and dynamics in simple flow situations. These aspects have been studied theoretically~\cite{sygulski2007stability,newman1991stability},
computationally~\cite{mavroyiakoumou2020large,mavroyiakoumou2021eigenmode,tiomkin2017stability,jaworski2012high,nardini2018reduced}, and experimentally~\cite{le1999unsteady}, and a wide range of membrane stability behavior and dynamics with various boundary conditions was revealed.
Some of these recent developments in membrane aerodynamics are reviewed in~\cite{lian2003membrane} and~\cite{tiomkin2021review}.

In~\cite{mavroyiakoumou2022membrane} we developed a mathematical model and numerical method to investigate the large-amplitude flutter of rectangular membranes that shed a trailing vortex-sheet wake in a three-dimensional (3-D) inviscid flow. There, we studied basic features of the membrane shapes and dynamics (steady, periodic, chaotic) and how they vary with the boundary conditions, fixed or free, at the membrane edges. Now, we build on that work and study in more detail the time-dependent spanwise variations in membrane deflection and how they depend on various material and geometric parameters. In~\cite{mavroyiakoumou2022membrane} we compared dynamics in a 2-D flow with those in a 3-D flow in the quasi-2-D limit (large span, free side-edges) and found good agreement. To explore this point further we now increase the spanwise resolution by a factor of four over that used in~\cite{mavroyiakoumou2022membrane}.
The main question that we answer in this work is: when are 3-D membrane flutter motions truly three-dimensional and when are they essentially 2-D? If the membrane shape does not vary much along the span, as seen in models of flapping flags in 3-D flows \cite{huang2010three}, a 2-D flow model may provide a good approximation to the motion, perhaps with a correction factor that depends on the aspect ratio (the ratio of the span to the chord length) \cite{eloy2007flutter}. If the membrane deflection does vary along the span, but in a simple way such as by a scaling factor that does not vary along the chord, other approximate 2-D flow models can be considered. For example,
the blade element theory applies fluid forces from 2-D flow models to each cross-section of a spinning propeller or turbine \cite{hansen2015aerodynamics,carlton2018marine}.
However, if the shape and dynamics vary strongly along the span, the 3-D model is more essential. In this work we determine when membrane dynamics vary strongly along the span, what form the dynamics take, and identify trends with respect to key parameters such as membrane mass, tension, Poisson ratio, and aspect ratio. We consider passive membrane deformations in this work, but the flows resulting from {\it prescribed} spanwise traveling-wave deformations have also been studied recently, as a method for reducing viscous drag in turbulent boundary layers \cite{du2002drag,choi2002drag,quadrio2004critical}.

Perhaps the simplest class of structures that show nontrivial dynamics along the span in fluid-structure interaction (FSI) is elastic cables, typically represented as 1-D curvilinear elastic rods in a 3-D flow \cite{ma2022flexible}. Important applications are the dynamics of mooring cables of submerged floating tunnels~\cite{ge2008nonlinear} under vortex-induced vibrations, resulting in spanwise waves~\cite{wu2012review}. A variety of phenomenological force-coefficient models \cite{facchinetti2004vortex} and direct numerical simulations~\cite{newman1997direct,evangelinos2000dns} have been used to compute the fluid forces on cables in uniform and shear flows. In both cases, standing- and traveling-wave dynamics are typical \cite{bourguet2011vortex}. Transitions between the two have been related to the cable mass density, tension, the distributions of hydrodynamic and structural damping, and the Reynolds number (Re)~\cite{newman1997direct,ma2022flexible}.

Like the present work, a number of recent studies have considered the dynamics of 2-D elastic objects in 3-D flows. A famous example of important spanwise dynamics is the torsional waves that led to the collapse of the original Tacoma Narrows Bridge~\cite{plaut2008snap,arioli2015new}. A particular area of interest is the dynamics of 2-D flexible flags or plates in 3-D flows. In \cite{huang2007simulation}, using the immersed-boundary method (IBM) and a uniform background flow, they found a flapping state that was slightly nonuniform along the span, but symmetric about the midspan cross-section, at $\mathrm{Re} = 100$--500. In \cite{tian2012onset,chen2020flapping}, similar, nearly spanwise-uniform motions were found in uniform and Poiseuille background flows respectively. In \cite{yu2012numerical}, a fictitious-domain FSI method with neo-Hookean elasticity for the flag was used to study the effect of aspect ratio at Reynolds numbers 100--800. They found flapping with spanwise-asymmetric warping motions at low aspect ratios ($\leq 0.6$), versus spanwise-symmetric flapping (but somewhat nonuniform along the span) at larger aspect ratios ($\geq 0.8$). A spanwise-periodic membrane flag in a 3-D viscous flow (Re $\approx 1000$) was studied by \cite{banerjee2015three}, with a small bending rigidity to avoid ill-posedness when tension in the flag vanishes. The membrane model included in-plane tension and shear stress like ours but with some differences. For example, the tension in \cite{banerjee2015three} is based on a 1-D stretching formula while ours results from a strain tensor given by deviations of the metric tensor from a reference metric tensor \cite{efrati2009elastic,alben2019semi}. 
The dynamics in~\cite{banerjee2015three} included standing waves along the span for a purely chordwise background flow (the case considered here), as well as oblique traveling waves with an oblique (chordwise and spanwise) background flow (not considered here). In \cite{jaiman2023isolated}, an arbitrary Lagrangian-Eulerian method for a finite-span flag was used at $\mathrm{Re} = 5000$, and they found a spanwise-symmetric flapping state like \cite{huang2007simulation,yu2012numerical} (i.e., with slightly larger flapping amplitude at the corners) as well as an asymmetric state with a slight bias towards one spanwise direction, depending on the flag mass and aspect ratio. 

The aforementioned studies considered viscous flows, while the present study considers 
2-D membranes in 3-D inviscid flows with vortex shedding from the trailing edge, which saves computational time.
A number of other studies have considered similar inviscid flow models but with flags of finite bending rigidity. In \cite{hiroaki2021numerical,hiroaki2021three,hiroaki2021theoretical} the vortex-lattice method was used with 11 panels in the spanwise direction (similar spanwise resolution to our previous membrane work \cite{mavroyiakoumou2022membrane}). They found flapping states that were essentially uniform along the span. The vortex-lattice method used here is a general approach that has been used more broadly to study 3-D inviscid flows past deformable surfaces with various geometries and boundary conditions~\cite{smyth2023generating,gibbs2012theory,gibbs2015stability,tang1999limit,tang2001effects,murua2012applications}.

The structure of the paper is as follows. \S\ref{sec:model} describes our model for the large-amplitude dynamics of a membrane in a 3-D inviscid flow with a vortex-sheet wake, over long times. The model is almost the same as in~\cite{mavroyiakoumou2022membrane} but with a more accurate elastic strain formula and higher spanwise resolution. We solve this nonlinear model using Broyden's method and an unsteady vortex lattice algorithm. \S\ref{sec:initialConditions} compares the membrane dynamics with two types of initial perturbations, in both the small- and large-amplitude regimes.
In \S\ref{sec:ffff} and \S\ref{sec:frfr}, we describe the computed membrane dynamics and how these vary with key parameters such as the membrane mass, membrane pretension, and Poisson ratio, for two sets of boundary conditions. We identify a variety of complex spanwise deformations in both cases, including oscillatory motions, periodic and nonperiodic in time, with and without spanwise symmetry, and with wavelike behavior in some cases. Multiple types of dynamics are often observed over long times within a single simulation. \S\ref{sec:aspectRatio} compares the dynamics of membranes with aspect ratios one, two, and four, and \S\ref{sec:spanAndChordPrestrain} studies the effect of anisotropic pretension (or pre-strain) in the membrane. \S\ref{sec:conclusions} gives the conclusions.

\begin{table}[H]
\centering
\begin{tabular}{|ll|}
\hline
\multicolumn{2}{|l|}{\textbf{Nomenclature}}\\ [.2cm]
$U\widehat{\mathbf{e}}_x$ & far field fluid velocity\\
$h$ & membrane thickness \\
$W$ & membrane spanwise width  \\
$2L$ & membrane chord\\
$\chi$  & aspect ratio ($\equiv W/(2L)$) \\
$\rho_s$ & mass per unit volume of the membrane \\
$\rho_f$ & mass per unit volume of the fluid\\
$\nu$ & Poisson ratio \\
$E$ & Young's modulus\\
$R_1$ & dimensionless membrane mass density ($=\rho_s h/(\rho_f L)$)\\
$R_3$ & dimensionless membrane stretching rigidity ($= Eh/(\rho_fU^2L)$)\\
$K_s$  & dimensionless membrane stretching stiffness ($=R_3/(1-\nu^2)$)\\
$\overline{e}_c, \overline{e}_s$  & pre-strains in the chordwise ($x$) and spanwise ($y$) directions\\
$\overline{e}$ & pre-strain in the uniform case ($\overline{e}\equiv\overline{e}_c=\overline{e}_s$)\\
$T_0$ & pretension \\
$T_{0c}$ & pretension in the chordwise direction\\
$T_{0s}$ & pretension in the spanwise direction \\
$\alpha_1$, $\alpha_2$ & membrane material coordinates ($x$ and~$y$ coordinates in the flat state)   \\
$[p](\alpha_1,\alpha_2,t)$ & fluid pressure jump ($=p_+(\alpha_1,\alpha_2,t)-p_-(\alpha_1,\alpha_2,t)$)   \\
F & fixed membrane edge \\
R & free membrane edge \\
FFFF & membranes with all four edges fixed \\
FRFR & membranes with the leading and trailing edges fixed and side edges free\\
\hline
\end{tabular}
\end{table}


\section{Large-amplitude membrane-vortex-sheet model}\label{sec:model}

\begin{figure}[ht]
    \centering
    \includegraphics[width=\textwidth]{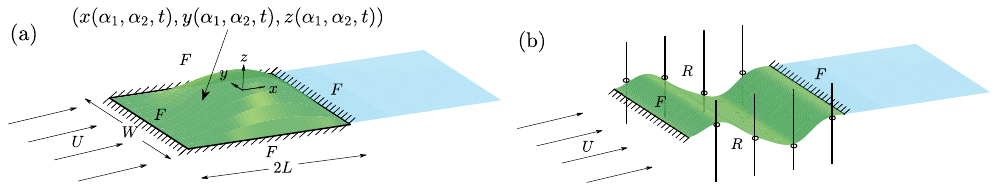}
    \caption{Schematic diagram (in perspective view) showing a 3-D membrane (dark green surface) with (a) all edges fixed (FFFF) and (b) fixed leading and trailing edges and free side-edges (FRFR). Along a free edge, points are fixed to massless rings that slide without friction along vertical poles. $U\widehat{\mathbf{e}}_x$ is the oncoming flow velocity, $W$ is the membrane's spanwise width, and $2L$ is the membrane's chord. There is also a flat vortex wake (light blue surface) that emanates from the membrane's trailing edge. The arrows indicate the far-field flow direction which is the same for each configuration.}
    \label{fig:schemBC}
\end{figure}

The membrane and flow models in this work are almost the same as in \cite{mavroyiakoumou2022membrane} but we repeat them briefly for completeness. 
We model the dynamics of an extensible membrane held in a 3-D fluid flow with velocity $U\widehat{\mathbf{e}}_x$ in the far field (see figure~\ref{fig:schemBC}). After a transient initial perturbation, the membrane edges are flat and parallel to the flow in the $z=0$ plane. The membrane surface is parametrized by two spatial coordinates and time, $\mathbf{r}(\alpha_1,\alpha_2,t)=(x(\alpha_1,\alpha_2,t),y(\alpha_1,\alpha_2,t),z(\alpha_1,\alpha_2,t))\in\mathbb{R}^3$, where the spatial coordinates $\alpha_1\in[-L,L]$ and $\alpha_2\in[-W/2,W/2]$ are material coordinates given by the $x$ and~$y$ coordinates of each point in the flat state. 

In the flat state, the membrane has in-plane strain corresponding to a pre-stretching from a zero-strain flat state. In this 
zero-strain or ``reference" state \cite{efrati2009elastic},
the membrane is a rectangle with dimensions smaller than
the $2L$-by-$W$ boundaries shown in figure~\ref{fig:schemBC}. It is stretched to make the $x$ and~$y$ coordinates of its edges match those of the boundaries, which we use to define its aspect ratio, $\chi \equiv W/(2L)$. This initial stretching is
assumed to be either constant and uniform (corresponding to strain~$\overline{e}$, the ``pre-strain") or biaxial (with constant pre-strains $\overline{e}_c$ in the chordwise ($x$) direction and $\overline{e}_s$ in the spanwise ($y$) direction). 

Each of the four membrane edges is either fixed at zero deflection or free to move in the $z$ direction, perpendicular to the oncoming flow. In the current work, we focus only on the FFFF and FRFR boundary conditions, where F stands for a \textit{fixed} edge and R stands for a \textit{free} edge. The first letter in each label is the leading-edge boundary condition type and the following letters are the boundary conditions moving clockwise around the rectangular membrane looking down from larger $z$ values, as in~\cite{mavroyiakoumou2022membrane}. Thus, the second and fourth letters are the side-edge boundary conditions and the third letter is the trailing-edge boundary condition. The equations for the two sets of boundary conditions are as follows:
\begin{align}
    \text{FFFF:}&\quad   z(-L,\alpha_2,t)=0,\;
    z(L,\alpha_2,t)=0,\; z(\alpha_1,-W/2,t)=0,\; z(\alpha_1,W/2,t)=0,\label{eq:bcffffn}\\
      \text{FRFR:}&\quad   z(-L,\alpha_2,t)=0,\;
    z(L,\alpha_2,t)=0,\;\partial_{\alpha_2} z(\alpha_1,-W/2,t)=0,\; \partial_{\alpha_2} z(\alpha_1,W/2,t)=0.\label{eq:bcfrfrn}
\end{align}
Whether the edge is fixed or free affects only the out-of-plane ($z$-) component of the membrane motion at the edge. In all cases, no in-plane motion of the membrane edges is allowed, i.e., the $x$- and $y$-coordinates of the edges are fixed:
\begin{align}
\mbox{Upstream/downstream edges:} \;
\left\{x\left(\pm L,\alpha_2,t\right)=\pm L \, , \; y\left(\pm L,\alpha_2,t\right)=\alpha_2\right\},\, & -W/2 \leq \alpha_2 \leq W/2,   \\ \mbox{Side edges:} \; \left\{x\left(\alpha_1,\pm W/2,t\right) = \alpha_1\,, \; y\left(\alpha_1,\pm W/2,t\right)= \pm W/2\right\},\, & -L \leq \alpha_1 \leq L.
\end{align}

The motivation for the free-edge boundary condition comes from our previous study~\cite{mavroyiakoumou2020large} of membrane dynamics in a 2-D flow, where the membrane is a 1-D curvilinear segment that undergoes large deflections. When both
membrane ends were held fixed at zero deflection, as in most previous studies of membrane flutter~\cite{le1999unsteady,sygulski2007stability,tiomkin2017stability,nardini2018reduced}, we found that all unstable membranes tended to a steady shape in the large-amplitude regime. If instead the membrane end is completely free to move, compression develops and the problem is ill-posed without a stabilizing effect such as bending rigidity, and difficult to solve computationally \cite{triantafyllou1994dynamic}. This motivated us to study membranes whose edges are free to move, but only in the direction perpendicular to the oncoming flow. This is equivalent to the free-end boundary condition for a string or membrane in classical mechanics~\cite{graff1975wave,farlow1993partial}, where the membrane end has horizontal slope. Physically, this `free' trailing edge means that the end slides without friction perpendicularly to the membrane's flat equilibrium state (for example, in~\cite{farlow1993partial} the end is attached to a frictionless, massless ring). Another way to simulate a free-end membrane condition, theoretically and experimentally, is by attaching the membrane edge to a thin rod or tether that can rotate freely about its other end at a pivot point~\cite{mavroyiakoumou2021dynamics,kashy1997transverse}. These works motivated our choice of boundary conditions in~\cite{mavroyiakoumou2022membrane} and in the present study.

The membrane dynamics are governed by the balance of momentum for a small material element with area $\Delta \alpha_1\Delta \alpha_2$: 
\begin{equation}
\rho_s h\Delta \alpha_1\Delta \alpha_2\partial_{tt}\mathbf{r}=\mathbf{f}_s\Delta \alpha_1\Delta \alpha_2-[p](\alpha_1,\alpha_2,t)\mathbf{\hat{n}}\sqrt{\det[\partial_{\alpha_i}\mathbf{r}\cdot\partial_{\alpha_j}\mathbf{r}]}\Delta \alpha_1\Delta \alpha_2,\label{eq:membrane3d}
\end{equation}
where $\rho_s$ is the mass per unit volume of the membrane, uniform in the undeformed state; $h$ is the membrane's thickness; $[p](\alpha_1,\alpha_2,t)$ is the fluid pressure; $\mathbf{\hat{n}}$ is the unit normal vector,
\begin{equation} \mathbf{\hat{n}}=(\partial_{\alpha_1}\mathbf{r}\times\partial_{\alpha_2}\mathbf{r})/\|\partial_{\alpha_1}\mathbf{r}\times\partial_{\alpha_2}\mathbf{r}\|,\label{eq:normal}
\end{equation} 
and $\det[\partial_{\alpha_i}\mathbf{r}\cdot\partial_{\alpha_j}\mathbf{r}]$ is the determinant of the metric tensor, so $\sqrt{\det[\partial_{\alpha_i}\mathbf{r}\cdot\partial_{\alpha_j}\mathbf{r}]}\Delta \alpha_1\Delta \alpha_2$ is the area of the material element in physical space. The stretching force per unit material area is given by 

    \begin{align}
\mathbf{f}_s=&\frac{Eh}{1+\nu} \left[\frac{\partial}{\partial \alpha_1}  \left(\frac{1}{1-\nu}(1+\bar{e}_c)^4 \epsilon_{11}\frac{\partial\mathbf{r}}{\partial \alpha_1}+\frac{\nu}{1-\nu}(1+\bar{e}_c)^2 (1+\bar{e}_s)^2\epsilon_{22}\frac{\partial\mathbf{r}}{\partial \alpha_1}+(1+\bar{e}_c)^2(1+\bar{e}_s)^2\epsilon_{12}\frac{\partial\mathbf{r}}{\partial \alpha_2} \right)\right.    \nonumber\\
&\hspace{.7cm} \left. +\frac{\partial}{\partial \alpha_2} \left( \frac{1}{1-\nu}(1+\bar{e}_s)^4 \epsilon_{22}\frac{\partial\mathbf{r}}{\partial \alpha_2}+\frac{\nu}{1-\nu}(1+\bar{e}_c)^2(1+\bar{e}_s)^2 \epsilon_{11}\frac{\partial\mathbf{r}}{\partial \alpha_2}+(1+\bar{e}_c)^2(1+\bar{e}_s)^2\epsilon_{12}\frac{\partial\mathbf{r}}{\partial \alpha_1}\right)  \right],\label{eq:3dtension}
\end{align}
and its derivation can be found in appendix~\ref{app:moreOnModel}. 
In~\eqref{eq:3dtension}, $E$ is Young's modulus, $\nu$ is Poisson's ratio, and 
\begin{equation}
    \epsilon_{11}=\frac{1}{2}\left(\partial_{\alpha_1}\mathbf{r}\cdot \partial_{\alpha_1}\mathbf{r}-\frac{1}{(1+\overline{e}_c)^2} \right); \quad 
    \epsilon_{12}=\frac{1}{2}\partial_{\alpha_1}\mathbf{r}\cdot \partial_{\alpha_2}\mathbf{r}; \quad 
    \epsilon_{22}=\frac{1}{2}\left(\partial_{\alpha_2}\mathbf{r}\cdot \partial_{\alpha_2}\mathbf{r}-\frac{1}{(1+\overline{e}_s)^2}\right), \label{eq:strainTensor}
\end{equation}
are the strain tensor components. Here the constants $\overline{e}_c$ and $\overline{e}_s$ are the aforementioned uniform pre-strain components (we do not use the small-$\overline{e}$ approximation of $\epsilon_{ij}$ from \cite{mavroyiakoumou2022membrane}). 
The biaxial pre-strain corresponds to a biaxial pre-tension of the membrane in the flat state:
\begin{align}
    T_{0c} &= \frac{K_s}{2}(1+\overline{e}_c)^2\left[\left((1+\overline{e}_c)^2-1\right)+\nu \left((1+\overline{e}_s)^2-1\right)\right],\label{eq:T0cformula}\\
    T_{0s}& =  \frac{K_s}{2}(1+\overline{e}_s)^2\left[\left((1+\overline{e}_s)^2-1\right)+\nu \left((1+\overline{e}_c)^2-1\right)\right].\label{eq:T0sformula}
\end{align}
In most cases we take $\overline{e}_c = \overline{e}_s \equiv \overline{e}$ in which case $T_{0c} = T_{0s} \equiv T_0$. We take $T_0$ as one of the main control parameters here, as in~\cite{mavroyiakoumou2020large,mavroyiakoumou2022membrane}. We briefly consider the case of
$\overline{e}_c \neq \overline{e}_s$ and $T_{0c} \neq T_{0s}$ in~\S\ref{sec:spanAndChordPrestrain}.

Equation~\eqref{eq:membrane3d} is made dimensionless by nondimensionalizing length by the membrane's half-chord $L$, time by $L/U$, and pressure by $\rho_f U^2$ where $U$ is the imposed fluid flow velocity and $\rho_f$ is the density of the fluid. The nonlinear, extensible membrane equation becomes
\begin{align}
     R_1{\partial_{tt}\mathbf{r}}-
K_s\left\{{\partial_{\alpha_1}} \left((1+\overline{e}_c)^4{\epsilon_{11}}{\partial_{\alpha_1}\mathbf{r}}+{\nu}(1+\overline{e}_c)^2(1+\overline{e}_s)^2{\epsilon_{22}}{\partial_{ \alpha_1}\mathbf{r}}+(1-\nu)(1+\overline{e}_c)^2(1+\overline{e}_s)^2{\epsilon_{12}}{\partial_{\alpha_2}\mathbf{r}} \right)\right.\nonumber\\
\left.+ {\partial_{ \alpha_2}}\left((1+\overline{e}_s)^4{\epsilon_{22}}{\partial_{\alpha_2}\mathbf{r}}+{\nu}{(1+\overline{e}_c)^2(1+\overline{e}_s)^2\epsilon_{11}}{\partial_{\alpha_2}\mathbf{r}}+(1-\nu)(1+\overline{e}_c)^2(1+\overline{e}_s)^2{\epsilon_{12}}{\partial_{\alpha_1}\mathbf{r}} \right)  \right\}\nonumber\\
    =-{[p]}\mathbf{\hat{n}}\sqrt{(\partial_{\alpha_1}\mathbf{r}\cdot \partial_{\alpha_1}\mathbf{r})(\partial_{\alpha_2}\mathbf{r}\cdot \partial_{\alpha_2}\mathbf{r})-(\partial_{\alpha_1}\mathbf{r}\cdot \partial_{\alpha_2}\mathbf{r})^2},\label{eq:nonlinearMembrane}
\end{align}
where $R_1=\rho_s h/(\rho_f L)$ is the dimensionless membrane mass density, $K_s=R_3/(1-\nu^2)$ is a dimensionless stretching stiffness written in terms of $R_3 = Eh/(\rho_fU^2L)$, the dimensionless stretching rigidity, and we have written out the determinant under the square root explicitly in \eqref{eq:nonlinearMembrane}. We have neglected bending rigidity, denoted $R_2$ in~\cite{alben2008flapping}. In the extensible membrane regime studied here, $R_3$ is finite, so $R_2 = R_3h^2/(12L^2)\to 0$ in the limit $h/L \to 0$.

We solve for the flow using the vortex lattice method, a type of panel method \cite{katz2001low} that solves for the 3-D inviscid flow past a thin body by posing a vortex sheet on the body to satisfy the no-flow-through or kinematic condition. The vortex sheet is advected into the fluid at the body's trailing edge, thus avoiding a flow singularity there \cite{katz2001low}. The
velocity $\mathbf{u}$ is a uniform background flow $\widehat{\mathbf{e}}_x$ plus the flow induced by a distribution of vorticity~$\bm{\omega}$, via
the Biot-Savart law \cite{saffman1992vortex}:
\begin{equation}\label{eq:BS}
    \mathbf{u}(\mathbf{x})=\widehat{\mathbf{e}}_x+\frac{1}{4\pi}\iiint_{\mathbb{R}^3}
    \bm{\omega}(\mathbf{x}',t) \times
    (\mathbf{x}-\mathbf{x}')/\|\mathbf{x} -\mathbf{x}'\|^3 \d\mathbf{x}'.
\end{equation}
\nn The vorticity is a vortex sheet on the body and wake. With the body surface parametrized by $\alpha_1$ and $\alpha_2$, we define a local coordinate basis by $\{\mathbf{\widehat{s}}_1, \mathbf{\widehat{s}}_2, \mathbf{\hat{n}}\}$:
\begin{equation}
    \mathbf{\widehat{s}}_1=\frac{\partial_{\alpha_1}\mathbf{r}}{\|\partial_{\alpha_1}\mathbf{r}\|};\qquad \mathbf{\widehat{s}}_2=\frac{\partial_{\alpha_2}\mathbf{r}}{\|\partial_{\alpha_2}\mathbf{r}\|};\qquad \mathbf{\hat{n}}\; \mbox{from (\ref{eq:normal})}.
\end{equation}
\nn Thus $\mathbf{\widehat{s}}_1$ and
$\mathbf{\widehat{s}}_2$ span the body's local tangent plane and $\mathbf{\hat{n}}$ is its normal vector.
$\mathbf{\widehat{s}}_1$ and $\mathbf{\widehat{s}}_2$ are also the tangents to the material lines $\alpha_2=\mathrm{constant}$ and $\alpha_1=\mathrm{constant}$, respectively. When the body experiences in-plane shear, 
$\mathbf{\widehat{s}}_1$ and $\mathbf{\widehat{s}}_2$ are not orthogonal, but they do not become parallel except for singular deformations that we do not consider. 

For the vortex sheet on the body, the vorticity takes the form $\bm{\omega}(\mathbf{x},t) = \bm{\gamma}(\alpha_1,\alpha_2,t)\delta(n) = \gamma_1(\alpha_1,\alpha_2,t)\delta(n)\mathbf{\widehat{s}}_1+\gamma_2(\alpha_1,\alpha_2,t)\delta(n) \mathbf{\widehat{s}}_2$, with $\delta(n)$ the Dirac delta distribution and $n$ the signed distance from the vortex sheet along the sheet normal. The vorticity is concentrated at the vortex sheet, $n = 0$, and $\bm{\gamma}$ is the jump in the tangential flow velocity across the vortex sheet \cite{saffman1992vortex}. The vorticity can be written similarly in the wake vortex sheet, but a different parametrization is used since $\alpha_1$ and
$\alpha_2$ are only defined on the body. The nonlinear kinematic equation states that the normal component of the body velocity equals that of the flow velocity at the body~\cite{saffman1992vortex}:
\begin{equation}\label{eq:kinematic}
    \mathbf{\hat{n}}\cdot \partial_t\mathbf{r}=\mathbf{\hat{n}}\cdot\left(\widehat{\mathbf{e}}_x+\frac{1}{4\pi}\Xiint{-}_{S_B+S_W}
    \bm{\gamma}(\mathbf{x}',t) \times
    (\mathbf{r}-\mathbf{x}')/\|\mathbf{r} -\mathbf{x}'\|^3 \d S_{\mathbf{x'}}\right),
\end{equation}
where $S_B$ and $S_W$ are the body and wake surfaces respectively. Since $\mathbf{r}$ and $\mathbf{x'}$ lie on $S_B$, the integral in (\ref{eq:kinematic}) is singular, defined as a principal value integral. 

The pressure jump across the membrane $[p](\alpha_1,\alpha_2,t)$ can be written in terms of the vortex sheet strength components $\gamma_1$ and $\gamma_2$ using the unsteady Bernoulli equation written at a fixed material point on the membrane. The formula and its derivation are given in~\cite{mavroyiakoumou2022membrane}, but the form of the pressure jump formula is
\begin{align}
\partial_{\alpha_{1}}[p] = G(\mathbf{r},\gamma_1,\gamma_2,\mu_1,\mu_2,\tau_1,\tau_2,\nu_v). \label{eq:pressureJumpAlpha1}
\end{align}
\nn Here $\mu_1$ and $\mu_2$ are 
the tangential components of the average of the flow velocity on the two sides of the membrane, i.e. the dot products of $\mathbf{\widehat{s}}_1$ and $\mathbf{\widehat{s}}_2$ with the term in parentheses on the right side of (\ref{eq:kinematic}). $\tau_1$, $\tau_2$, and $\nu_v$ are the components of the membrane's velocity in the $\{\mathbf{\widehat{s}}_1, \mathbf{\widehat{s}}_2, \mathbf{\hat{n}}\}$ basis:
\begin{equation}\label{eq:tau1}
\tau_1(\alpha_1,\alpha_2,t)=\partial_t\mathbf{r}\cdot \mathbf{\widehat{s}}_1;\qquad
\tau_2(\alpha_1,\alpha_2,t)=\partial_t\mathbf{r}\cdot \mathbf{\widehat{s}}_2;
\qquad\nu_v(\alpha_1,\alpha_2,t)=\partial_t\mathbf{r}\cdot\mathbf{\hat{n}}.
\end{equation}
We integrate it from the trailing edge using the Kutta condition
\begin{equation}
    [p](\alpha_1=1,\alpha_2,t)=0,
\end{equation}
to obtain $[p](\alpha_1,\alpha_2,t)$ at all points on the membrane.

In summary, equations (\ref{eq:nonlinearMembrane}), (\ref{eq:kinematic}), and (\ref{eq:pressureJumpAlpha1}) are a coupled system of equations for $\mathbf{r}$, 
$\bm{\gamma}$, and $[p]$ that we can solve with suitable initial and boundary conditions to compute the membrane dynamics.

In the current work, we use two different types of initial perturbations, one asymmetric and the other symmetric along the span. The membrane shape is prescribed at the first three time steps ($t = 0.05$, 0.1, and 0.15) and then the dynamical equations~\eqref{eq:nonlinearMembrane}, \eqref{eq:kinematic}, and \eqref{eq:pressureJumpAlpha1} give the membrane shape at $t = 0.2$ and the following time steps (uniformly spaced by 0.05).

The spanwise-asymmetric perturbation takes the form of a membrane that starts in the horizontal state at $t=0$ and then the edges are given a small transient skewed deformation, shown in the first snapshots of figures~\ref{fig:smallTransientLargeFFFF}(a) and (c) ($t = 0.05$), that smoothly grows and then decays back to the flat state over a time scale $\eta = 0.2$:
\begin{subequations}
\begin{align}
    \mathrm{FFFF:}\quad &z(\pm 1,\alpha_2,t)= \pm 0.1 \alpha_2\left(t/\eta\right)^3e^{-(t/\eta)^3},\quad  -\chi\leq \alpha_2\leq \chi,\label{eq:saddleShapePert}\\
   &z\left(\alpha_1,\pm \chi,t\right)=\pm 0.1\alpha_1\chi\left(t/\eta \right)^3e^{-(t/\eta)^3},\quad -1\leq \alpha_1\leq 1,\label{eq:saddleShapePert2}
\end{align}
\end{subequations}
for $t\geq 0$. The skewed perturbation for FRFR membranes is the same as~\eqref{eq:saddleShapePert} but is only applied to the fixed (upstream and downstream) edges, as shown in the first snapshots of figures~\ref{fig:smallTransientLargeFRFR}(a) and (c) ($t = 0.05$). For this first perturbation, the far-field oncoming flow is set to $U\left(1- e^{-(t/\eta)^3}\right)\widehat{\mathbf{e}}_x$, with $\eta$ the same as above. Thus the far-field flow smoothly transitions from rest to the long-time limit $U\widehat{\mathbf{e}}_x$. 

The spanwise-symmetric perturbation has a small nonzero slope in the interior, initially:
\begin{equation}\label{eq:inclinedPlaneIC}
    z(\alpha_1,\alpha_2,0) = \sigma \alpha_1,\quad -1< \alpha_1< 1,\quad  -\chi < \alpha_2 < \chi,
\end{equation}
with $\sigma=10^{-3}$, and $z = 0$ on the fixed edges.
This initial condition is shown in the first snapshots of figures~\ref{fig:smallTransientLargeFFFF}(b) and (d) and~\ref{fig:smallTransientLargeFRFR}(b) and (d) for FFFF and FRFR, respectively.

For some $(R_1,T_0)$ values, as $t$ exceeds 1 the membrane deflection decays exponentially in time and tends to the flat, horizontal state: a stable equilibrium. For other parameter values, the membrane deflection grows rapidly with time (sometimes exponentially), in which case the flat state is unstable. In the following sections, we show the large-amplitude dynamics of the membrane and how it depends on the initial and boundary conditions and the physical parameters. 

\section{Effect of initial conditions}\label{sec:initialConditions}

\begin{figure}[H]
    \centering
    \includegraphics[width=\textwidth]{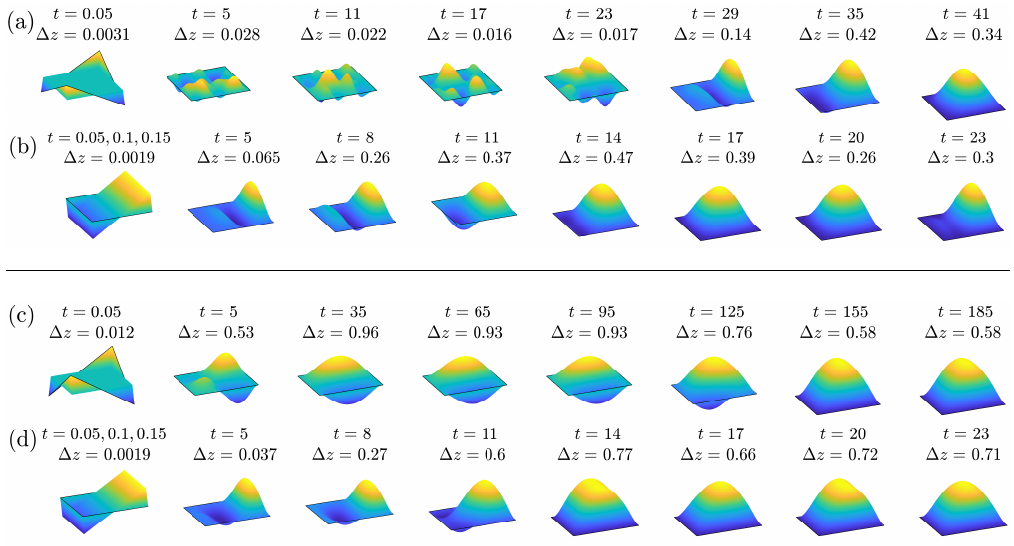}
    \caption{
        (FFFF) Each panel shows a sequence of eight 3-D snapshots, from left to right with increasing time. At the top of each snapshot the values of $t$ and $\Delta z=z_{\max}-z_{\min}$ are listed. Panels (a,c) result from the skewed initial condition~\eqref{eq:saddleShapePert}--\eqref{eq:saddleShapePert2} and panels (b,d) result from the spanwise-symmetric initial condition~\eqref{eq:inclinedPlaneIC}. Panels (a,b) are aspect-ratio-one membranes with $\nu =0$ and $(R_1,T_0) = (10^{-0.5},10^{-0.5})$. The motions in panels (a) and (b) are shown in the supplementary movies named ``Fig2a\dots mp4" and ``Fig2b\dots mp4,'' respectively. Panels (c,d) show aspect-ratio-four membranes with $\nu=0.5$ and $(R_1,T_0)=(10^{-0.5},10^{-0.3})$.}\label{fig:smallTransientLargeFFFF}
\end{figure}

\indent\indent We begin by considering the effect of the initial conditions on the membrane flutter dynamics. For the numerical results we use one value of the stretching rigidity, $R_3=1$, that gives relatively large deflections during flutter. In figure~\ref{fig:smallTransientLargeFFFF} we show four snapshot sequences, grouped into two pairs ((a)--(b) and (c)--(d)) that illustrate the effect of the initial conditions on the transition from small- to large-amplitude deflections in the FFFF case. The first sequence in each pair (panels (a) and (c)) has the skewed initial condition, while the second sequence has the symmetric initial condition (panels (b) and (d)). Otherwise, the physical parameters are the same within each pair: (a)--(b) have aspect ratio one, $\nu=0$, and 
$(R_1,T_0)=(10^{-0.5},10^{-0.5})$; (c)--(d) have 
aspect ratio four, $\nu=0.5$, and $(R_1,T_0)=(10^{-0.5}, 10^{-0.3})$.
The motions in figures~\ref{fig:smallTransientLargeFFFF}(a) and (b) are shown in the supplementary movies named ``Fig2a\dots mp4" and ``Fig2b\dots mp4,'' respectively.

In panel (a), the skewed perturbation leads immediately to a spanwise-asymmetric motion with many small peaks and troughs whose positions change rapidly in time (shown at $t=5$, 11, 17, and 23). Near $t = 29$, the membrane suddenly shifts to a spanwise-symmetric shape with three extrema along the chordwise direction: a small peak and trough near the leading edge and a large peak near the trailing edge. By $t = 35$ the small peak near the leading edge has disappeared and the trough has moved forward, and by $t = 41$, the membrane has assumed a single-hump shape that remains spanwise-symmetric at later times. In panel (b), the spanwise-symmetric initial condition instead leads directly to the three-extrema shape by $t = 5$, which is followed by a similar sequence of transformations leading to a single hump shape. Pair (a)--(b) is one example of many cases where the different initial conditions eventually lead to the same state.

In panels (c)--(d), the different initial conditions lead to different motions for over 100 time units, but in this case (and most others) the membranes eventually reach the same steady state, which in this case is a steady single hump shape.
The transitional dynamics for the skewed perturbation in panel (c), like in (a), at first show a complex and asymmetric distribution of small transient peaks and troughs with shifting positions and amplitudes. By $t = 35$, the membrane assumes a smooth and nearly steady shape with one peak and one trough along the span, antisymmetric about the midspan cross-section. This shape is maintained until shortly before $t = 155$, when the trough grows and peak diminishes rapidly, leaving a single-hump shape. The motion in panel (d) reaches a steady single-hump shape much earlier than in panel (c), and the transitional dynamics resemble those in panel (b). In general, the initial condition that matches the steady-state motion in terms of the presence or absence of spanwise symmetry usually reaches the steady-state motion much faster. The other initial condition often attains a metastable state for long times before abruptly transitioning to the steady state, as in panel (c).

\begin{figure}[H]
    \centering
    \includegraphics[width=\textwidth]{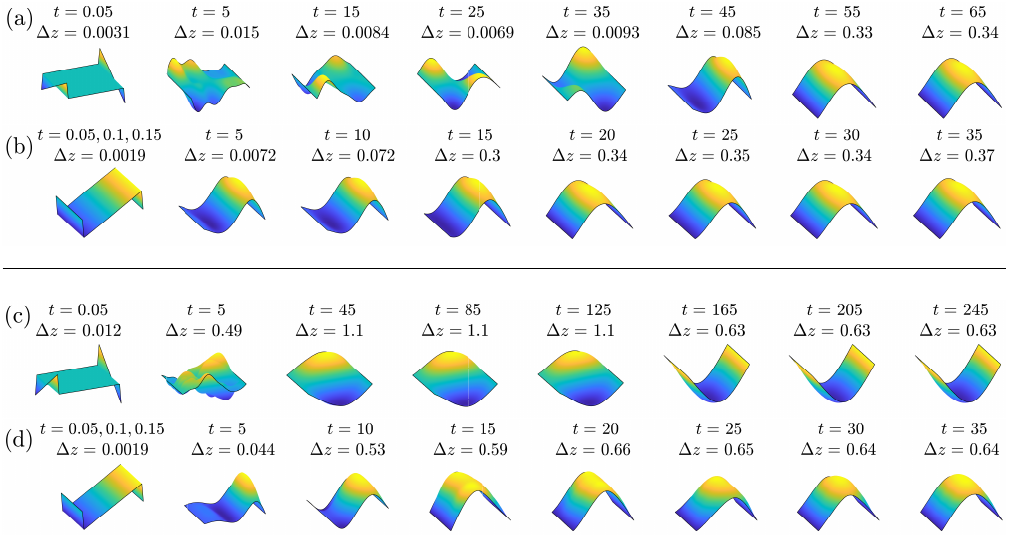}
    \caption{Same as figure~\ref{fig:smallTransientLargeFFFF}  but for FRFR membranes and panels (a,b) for aspect-ratio-one membranes with $\nu =0$ and $(R_1,T_0) = (10^{-1.5},10^{-0.25})$, panels (c,d) for aspect-ratio-four membranes with $\nu=0.5$ and $(R_1,T_0)=(10^{0},10^{-0.375})$. The motions in panels (a) and (b) are shown in the supplementary movies named ``Fig3a\dots mp4" and ``Fig3b\dots mp4,'' respectively.}
     \label{fig:smallTransientLargeFRFR}
\end{figure}

In figure~\ref{fig:smallTransientLargeFRFR} we show the effect of the initial conditions on the transition to large-amplitude dynamics in two cases with FRFR boundary conditions instead of FFFF. Panels (a) and (b) show the comparison for aspect ratio one, $\nu=0$, and $(R_1,T_0)=(10^{-1.5},10^{-0.25})$. The motions in these panels are shown in the supplementary movies named ``Fig3a\dots mp4" and ``Fig3b\dots mp4,'' respectively. In panel (a), with the asymmetric initial condition~\eqref{eq:saddleShapePert}, the membrane first exhibits strong spanwise asymmetry, then becomes almost spanwise symmetric by $t=45$, and maintains this spanwise symmetry until the end of the simulation at $t\approx 1000$.
Figure~\ref{fig:smallTransientLargeFRFR}(b) shows that the symmetric initial condition  instead grows in deflection amplitude as a spanwise-symmetric shape with two extrema at early times and then reaches a single-hump shape like that in panel (a) but somewhat earlier ($t \approx $ 20). Similar to the long-time dynamics in panel (a), it maintains spanwise symmetry until the end of the simulation.

Panels (c)--(d) compare the two initial conditions at aspect ratio four, $\nu=0.5$, and $(R_1,T_0)=(10^{0.5},10^{-0.5})$. Here, similar to figures~\ref{fig:smallTransientLargeFFFF}(c) and (d) for FFFF membranes, the asymmetric initial perturbation in panel (a) leads to complicated wavy motions before it becomes a two-bump shape in the large-amplitude regime at $t\in[45,125]$---with the peak and trough along the span. It finally transitions to a motion with unsteady spanwise-asymmetric fluctuations about a single-hump shape. The symmetric initial condition (d) remains symmetric at early times, reaching a single-hump shape much earlier ($t\approx 15$ versus 165 in panel~(c)). It then becomes unsteady for some time but remains nearly symmetric, and at later times it becomes steady.
As in figure~\ref{fig:smallTransientLargeFFFF}(c)--(d), both initial conditions tend toward the same long-time steady states, but approach different metastable states at earlier times that reflect the spanwise symmetry/asymmetry of the initial conditions.

\begin{figure}[t]
    \centering
    \includegraphics[width=\textwidth]{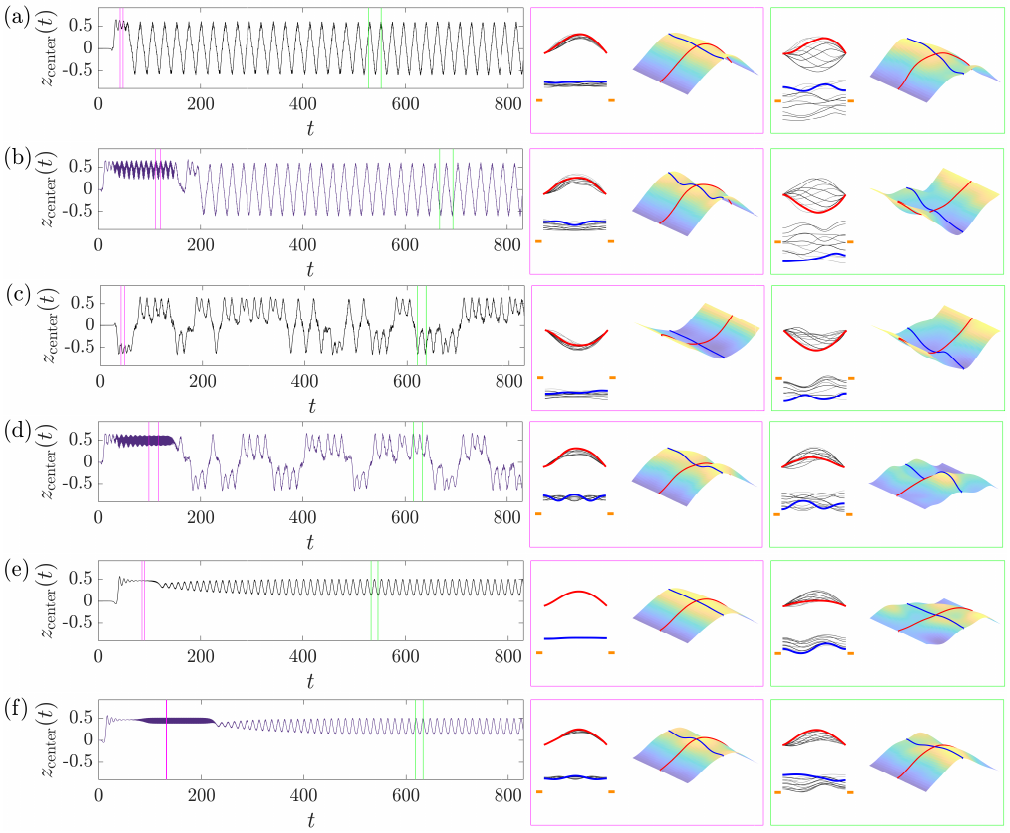}
    \caption{Dynamics of FRFR membranes with aspect ratio one, $\nu=0.5$, in six cases: symmetric and asymmetric initial conditions at three pairs of $(R_1,T_0)$ values. At the left of each panel, $z_{\mathrm{center}}(t)$ is plotted, indicating changes in dynamics over long times. Further to the right (in the second and third columns), sets of 11 membrane midspan and midchord snapshots are shown in the magenta and green boxes respectively. The snapshots are equally spaced in time over intervals bounded by the magenta and green vertical lines in the $z_{\mathrm{center}}(t)$ plots, respectively.
    Panels (a,b) correspond to $(R_1,T_0)=(10^{-1.5},10^{-0.5})$ with the asymmetric initial condition~\eqref{eq:saddleShapePert} and the symmetric initial condition~\eqref{eq:inclinedPlaneIC}, respectively. Panels (c,d) have $(R_1,T_0)=(10^{-1},10^{-0.5})$ and panels (e,f) have $(R_1,T_0)=(10^{-1.5},10^{-0.375})$, with the same initial conditions. The $z=0$ plane is marked by the small orange lines for the midchord snapshots (and by the leading and trailing edges for the midspan snapshots). The motions in panels (b) and (f) are shown in the supplementary movies named ``Fig4b\dots mp4" and ``Fig4f\dots mp4,'' respectively. 
    }\label{fig:FRFRwithAR1saddleAndLinearIC}
\end{figure}
\FloatBarrier
In figures~\ref{fig:smallTransientLargeFFFF} and \ref{fig:smallTransientLargeFRFR} the focus was on the transition from small to large deflections for the two different initial conditions. Figure~\ref{fig:FRFRwithAR1saddleAndLinearIC} shows the effect of the initial conditions on the large-amplitude dynamics only, over longer times. We consider FRFR aspect-ratio-one membranes with $\nu=0.5$ and three pairs of $(R_1,T_0)$, as listed in the caption, one for each pair of panels (a)--(b), (c)--(d), and (e)--(f), with the asymmetric initial condition~\eqref{eq:saddleShapePert} used for the first of each pair ((a), (c), and (e)), and the symmetric initial condition~\eqref{eq:inclinedPlaneIC} used for the second ((b), (d), and (f)). The motions in panels (b) and (f) are shown in the supplementary movies named ``Fig4b\dots mp4" and ``Fig4f\dots mp4,'' respectively.  In each panel in figure~\ref{fig:FRFRwithAR1saddleAndLinearIC} we show on the left $z_{\mathrm{center}}(t)$, the $z$-deflection of the membrane center versus time. On the right, within each of the magenta and green boxes, we present sequences of membrane snapshots that illustrate early- and late-time dynamics during $t\in[0,850]$, in time intervals marked by the magenta and green lines on the
$z_{\mathrm{center}}(t)$ plots. The snapshots are sequences of lines, ranging from light gray at earlier times to dark gray at later times, that end with either a red or blue line. The set ending with the red line is the midspan snapshots---they show the membrane position along the cross-section in the chordwise direction, midway across the span. The set ending with the blue line is the midchord snapshots---the membrane position along the cross-section in the spanwise direction, midway across the chord. Next to the sequences of midspan and midchord snapshots, a single 3-D surface plot is shown at the last time in the sequences, with red and blue lines showing the last midspan and midchord snapshots respectively. These two snapshot sequences present only a subset of the information in the full membrane motion, but they illustrate the motion using relatively little space, so we use them extensively to illustrate how motions vary with different parameters and boundary conditions.

In all the $z_{\mathrm{center}}(t)$ plots in figure~\ref{fig:FRFRwithAR1saddleAndLinearIC}, $z_{\mathrm{center}}(t)$ is very small over an early time interval in which the initial perturbation grows as shown in figures \ref{fig:smallTransientLargeFFFF} and 
\ref{fig:smallTransientLargeFRFR}. The asymmetric perturbation ((a), (c), and (e)) reaches large amplitude later than the symmetric perturbation ((b), (d), and (f)) in each case. The membrane deflections first reach large amplitudes with dynamics (shown in the magenta boxes) that differ from the eventual steady states (shown in the green boxes). Panels (a) and (b) first reach a state where there are oscillations about an upward-deflected state, larger in panel (b) than in panel (a), as shown in the magenta boxes. This metastable state persists longer in (b) ($\approx 180$ time units), but both motions eventually transition to the same spanwise-asymmetric periodic up-and-down oscillation, shown in the green boxes. Panels (c) and (d), with a slightly larger mass ($R_1=10^{-1}$), show a similar transition from different metastable states (magenta boxes)
to essentially the same large-amplitude state (green boxes), which is nonperiodic now.
Panels (e) and (f), with smaller $T_0$ ($10^{-0.375}$) show a somewhat different sequence of states.
Both (e) and (f) first reach a nearly-steady single-hump shape, and then panel~(f) transitions to a periodic oscillation, shown in the magenta box, that does not occur in (e). Eventually both membranes reach the same spanwise-asymmetric upward-deflected periodic oscillations.
The panels (a)--(f) have certain common features that are seen more widely in parameter space. The membranes tend to oscillate about a primarily upward-deflected state, a primarily downward-deflected state, or between the two. In all of these panels, the membranes initially reach a primarily upward- or downward-deflected state before transitioning to the eventual steady-state motion. Panels (a) and (b) assume primarily up-and-down oscillations, panels (c) and (d) transition between primarily upward and downward states irregularly at large times, while (e)--(f) assume upward-deflected oscillations. All the panels show oscillations with similar dominant frequencies because $R_1$ is small. Superposed on the dominant oscillations are much-higher-frequency spanwise-asymmetric oscillations that can be seen in the slight jaggedness of the plots in panels (c) and (d) but also occur with even finer jaggedness in (a)--(b), (e)--(f). The main message of figure~\ref{fig:FRFRwithAR1saddleAndLinearIC} is that the same long-time steady-state dynamics apparently occur with both initial conditions, but there are long transient times ($\approx 200$--$300$ time units) in which the dynamics may differ greatly. The transient time may in fact be much longer when $R_1 \gg 1$.

\section{All edges fixed}\label{sec:ffff}

We have compared the transient and long-time dynamics in several cases with both symmetric and asymmetric initial conditions. Now we characterize more fully the typical dynamics of membranes with all edges fixed. We present the membrane snapshots only for the asymmetric initial condition henceforth, but compare the time evolution of the spanwise asymmetry for both initial conditions.
In~\cite{mavroyiakoumou2022membrane} we described three types of dynamics for FFFF membranes, one steady and two unsteady. Now,
using 40 instead of 10 spanwise panels, we describe the spanwise variations in the dynamics and their time evolution, which was not discussed in \cite{mavroyiakoumou2022membrane} but is the main focus here. We also use a strain formula \eqref{eq:strainTensor} that is more accurate for large deflections, and consider effects of Poisson ratio, and (in later sections) aspect ratio and anisotropic pretension.
We focus exclusively on the stretching stiffness $R_3 = 1$, which is small enough that a wide variety of large-amplitude dynamics are seen, but large enough that the simulations generally converge at every time step for long times, and the membrane deflections are not unrealistically large.

\begin{figure}
    \centering
    \includegraphics[width=.85\textwidth]{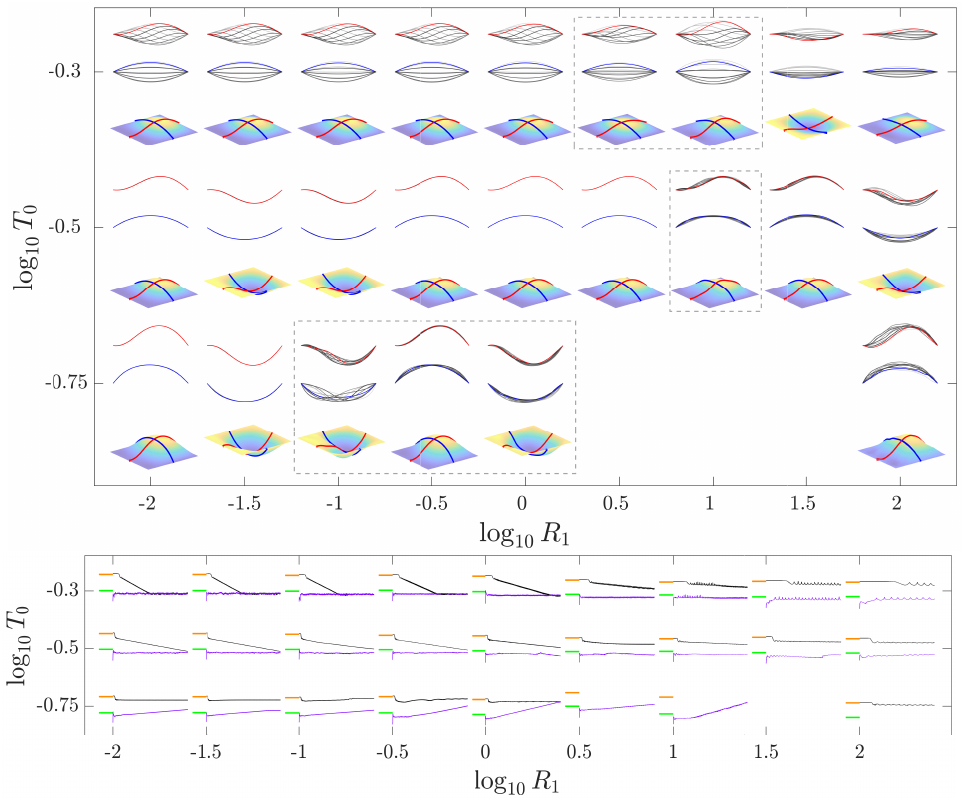}
    \caption{Snapshots of FFFF membranes at large times over
    $R_1$-$T_0$ space. At each ($R_1$, $T_0$) pair, three sets of snapshots are shown, from top to bottom: midspan snapshots, midchord snapshots, and one 3-D membrane surface snapshot with the midspan and midchord snapshots highlighted in red and blue. The membranes have aspect ratio one and $\nu=0$. Snapshots are omitted in the lower-right corner, at $R_1\in [10^{0.5}, 10^{1.5}]$ and $T_0=10^{-0.75}$, because large-amplitude, steady-state membrane motions were not obtained. The large panel at the bottom shows $\log_{10}|z_{\mathrm{asymm}}(t)|$ over the same portion of $R_1$-$T_0$ space. The plots are given in pairs, black and purple for the asymmetric and symmetric initial conditions, respectively. At the left end and near the bottom of each pair, the 
    $R_1$ and $T_0$ values are listed on the corresponding axes. The 
    horizontal $t$ and vertical $\log_{10}|z_{\mathrm{asymm}}|$ axes are omitted, but
    for each pair of plots, the green and orange hash marks denote $\log_{10}|z_{\mathrm{asymm}}|= -7$ and 0 respectively.}    \label{fig:FFFFar1nu0}
\end{figure}

As in \cite{mavroyiakoumou2022membrane}, we identify the stability boundary in $R_1$-$T_0$ space by applying the small initial perturbations in figures \ref{fig:smallTransientLargeFFFF} and \ref{fig:smallTransientLargeFRFR} across a wide range of values and identifying where the perturbations grow or decay exponentially in time. In appendix \ref{app:StabilityBoundaries} we show that the exponential growth has an oscillatory part (divergence with flutter) slightly inside the instability region, but further inside the instability region we have divergence without flutter. Within the instability region for FFFF membranes with aspect ratio one and $\nu$ = 0, figure~\ref{fig:FFFFar1nu0} shows sequences of midspan and midchord snapshots across a wide range of $R_1$ and $T_0$ values.
The types of dynamics are similar to those with the linearized strain formula in 
\cite{mavroyiakoumou2022membrane}, as well as their ordering with $T_0$. In particular, at the largest $T_0$, 10$^{-0.3}$ in figure~\ref{fig:FFFFar1nu0} (as well as 10$^{-0.25}$, not shown), the membrane performs an up-down oscillation with traveling waves that move upstream (i.e., leftward for the midspan snapshots, those in the top of the three rows at each ($R_1$,~$T_0$) pair). 
At the two largest $R_1$, a smaller oscillation is seen instead. 
Moving to the row below, $T_0 = 10^{-0.5}$, the membranes are steady or perform spanwise symmetric oscillations. At $T_0=10^{-0.75}$, the membranes are steady or perform spanwise-asymmetric oscillations about primarily upward or downward mean deflections. The transition from up-down oscillations to primarily upward- or downward-deflected states was also seen in \cite[figure 12]{mavroyiakoumou2022membrane}. Here and in most other cases, the motions are generally more irregular and spanwise-asymmetric at smaller $T_0$ and larger $R_1$.

We now define a measure of the spanwise asymmetry of the membrane shape at each time:
\begin{align}\label{eq:zasymmetry}
    z_{\mathrm{asymm}}(t) &\equiv \int_{-\chi}^\chi\int_{-1}^1 |z(\alpha_1,\alpha_2,t)-z_{\mathrm{refl}}(\alpha_1,\alpha_2,t)|\d \alpha_1\d \alpha_2\bigg/\int_{-\chi}^{\chi}\int_{-1}^1 |z(\alpha_1,\alpha_2,t)|\d\alpha_1\d\alpha_2, \\ z_{\mathrm{refl}}(\alpha_1,\alpha_2,t) &\equiv z(\alpha_1,-\alpha_2,t).
\end{align}
Here $z_{\mathrm{refl}}$ is $z$ for the membrane shape reflected in the plane $\alpha_2 = 0$ (i.e., the midspan plane), so
$z_{\mathrm{asymm}}(t) = 0$ when the membrane deflection is symmetric with respect to the midspan.

At the bottom of figure~\ref{fig:FFFFar1nu0}, the large panel shows $\log_{10}|z_{\mathrm{asymm}}(t)|$ plots in the same $R_1$-$T_0$ space as the snapshots above, for the symmetric (purple) and asymmetric (black) initial conditions. The axes are omitted to simplify the figure, but the plots run up to $t \approx 1000$ in each case and the green and orange hash marks denote  
$\log_{10}|z_{\mathrm{asymm}}| = -7$ and 0 respectively. 
At $T_0=10^{-0.3}$ and $10^{-0.5}$, and small-to-moderate values of $R_1$, $z_{\mathrm{asymm}}(t)$ generally decreases for the asymmetric initial condition and stays small for the symmetric initial condition throughout the time shown. In these cases a symmetric oscillation or steady deflected shape occurs. Moving to $T_0 = 10^{-0.75}$, $z_{\mathrm{asymm}}(t)$ grows slowly for the symmetric initial condition and oscillates at large times for the asymmetric initial condition. 
At $R_1 > 10^{1.5}$, $z_{\mathrm{asymm}}(t)$ generally grows at large times, though the growth rate decreases with increasing $R_1$. The large-time steady-state behavior may only be reached for $t \gg 1000$ at these $R_1$ values. Taken together, the plots show how the long-time dynamics vary over parameter space and initial conditions with respect to spanwise asymmetry, and show typical time scales needed to reach the long-time dynamics. At most ($R_1$, $T_0$) values, the two $z_{\mathrm{asymm}}(t)$ plots with different initial conditions show similar long-time behavior, but in a few cases, such as the two rightmost columns in the top row, there is little indication of eventual convergence.

\begin{figure}
    \centering
    \includegraphics[width=.97\textwidth]{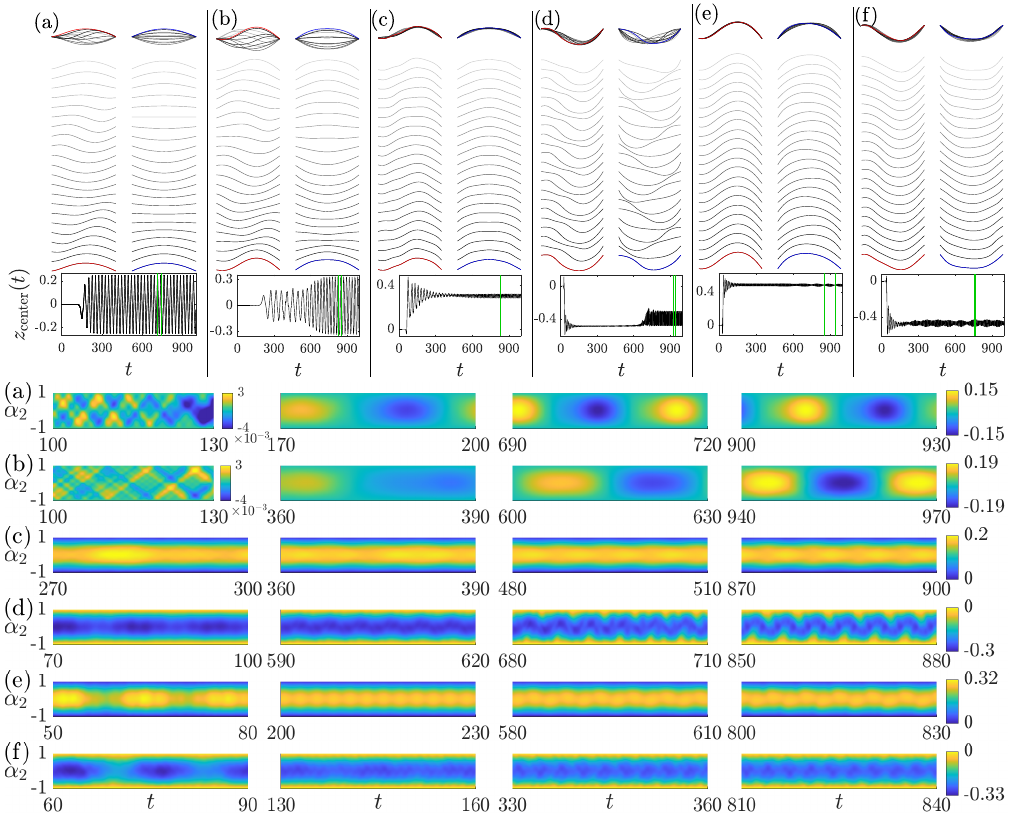}
\caption{Six examples of long-time dynamics for FFFF membranes with aspect ratio one and $\nu=0$. Panels (a)--(f) show sequences of midspan (left column) and midchord (right column) snapshots, overlaid at the top of each column, and arrayed vertically below, from top to bottom with increasing time, in darkening shades from gray followed by red and blue at the last times. The snapshots are taken between two relatively late times when the membrane center deflection $z_{\mathrm{center}}$ reaches successive upward or downward extrema. The $(R_1,T_0)$ values are (a) $(10^{0.5},10^{-0.3})$, (b) $(10^{1},10^{-0.3})$, (c) $(10^1,10^{-0.5})$, (d) $(10^{-1},10^{-0.75})$, (e) $(10^{-0.5},10^{-0.75})$, and (f) $(10^{0},10^{-0.75})$. Below the snapshots, $z_{\mathrm{center}}(t)$ is plotted, with the two green vertical lines marking the time range of the snapshots. At the bottom of the figure the color plots show $\langle z \rangle_c(\alpha_2, t)$, the chordwise average of $z$, over selected time intervals labeled at the bottom. The motions in panels (a)--(f) are shown in the supplementary movies named ``Fig6a\dots mp4" through ``Fig6f\dots mp4,'' respectively.}
    \label{fig:ffffar1nu0vertical}
\end{figure}

In figure~\ref{fig:FFFFar1nu0} we have indicated how the dynamics vary over $R_1$-$T_0$ space for a representative time interval between peaks of deflection at late times.  Now, in figure~\ref{fig:ffffar1nu0vertical}, we study six cases in more detail. We present the midspan and midchord dynamics more clearly and show the long-time dynamics that lead from the initial conditions to the quasi-steady-state behavior depicted by the snapshot sequences. In these long-time dynamics we highlight the membrane shape variations along the spanwise coordinate. We choose three values of $T_0$ ($10^{-0.3}$, $10^{-0.5}$, and $10^{-0.75}$) and five values of $R_1$ (listed in the caption), the cases in the gray dashed boxes in figure~\ref{fig:FFFFar1nu0}.

At the top of figure~\ref{fig:ffffar1nu0vertical} we array the midspan and midchord snapshots vertically, in left and right columns respectively, moving downward with increasing time for each case (a)--(f). The full set of snapshots are shown overlaid in physical space at the top of each column. Below the columns of snapshots, we plot $z_{\mathrm{center}}(t)$ in each case, with the time interval of the snapshots marked by the two green vertical lines (very closely spaced). In the bottom portion of the figure color plots are used to show the chordwise average of $z$ over time,
\begin{equation}\label{eq:chordwiseAvgInt}
    \langle z\rangle_c(\alpha_2, t) \equiv\frac{1}{2}\int_{-1}^1 z(\alpha_1,\alpha_2,t)\,\d\alpha_1,\quad -\chi\leq \alpha_2\leq \chi.
\end{equation}
$\langle z\rangle_c(\alpha_2, t)$ is a projection of $z(\alpha_1,\alpha_2,t)$ onto the two-dimensional $(\alpha_2, t)$ space so motions along the spanwise $\alpha_2$ coordinate are easier to visualize and chordwise motions are suppressed. In each case $\langle z\rangle_c(\alpha_2, t)$ is shown over four time intervals that highlight different stages in the dynamics.

The $z_{\mathrm{center}}(t)$ plots below the vertically arrayed snapshots are a guide to the overall dynamics in each case. After an initial period of growth from the flat state (that is longer for larger $R_1$ and larger $T_0$), the membranes show a large-amplitude motion that is depicted by the second of the four color plots in cases (a)--(b) and by the first of the four color plots in cases (c)--(f). In panels (a)--(b) the $z_{\mathrm{center}}(t)$ plots show that the dynamics reach large amplitude slowly---after 150--200 time units. During the initial period of growth, the membranes have waves in the spanwise direction with wave crests that move from the four edges inward, reflect off of each other, and move back toward the boundaries. 
For larger $R_1$ $(10^{1})$ in (b), the wave crests have smaller speed, shown by the smaller slopes of the diagonal bands. Panels (a) and (b) show examples of up-down oscillations that occur when $T_0 = 10^{-0.3}$. When the large-amplitude regime is reached in these cases, the oscillation frequency of the up-down motions increase with time. This is seen through the $z_{\mathrm{center}}(t)$ plots but also through the color plots in cases (a)--(b) at the bottom of figure~\ref{fig:ffffar1nu0vertical}, where going from the second to the third to the fourth plot the widths of the yellow (upward) and blue (downward) bumps decrease with time. The $z_{\mathrm{center}}(t)$ plots below the vertically arrayed snapshots in (b), (d), and (e), show that the large-amplitude dynamics change from one type to another after hundreds of time units. The change is relatively abrupt in panel (d), near $t = 700$, and more gradual in (b) and (e).
In panel (c), the first color plot shows a motion that oscillates about a spanwise symmetric single-hump shape that becomes almost steady in the second color plot. The third and fourth color plots show a transition to a motion with periodic modulations that involve alternate sharpening and flattening of the peak in the spanwise direction. The spanwise symmetric sharpening and flattening of panel (c) can be seen clearly in the midchord snapshot sequence (right column of panel (c) at the top of the figure). The corresponding midspan snapshots (left column of panel (c)) sharpen when the midchord snapshots flatten and vice versa.  Therefore, in 3-D the peak alternately spreads in the chordwise and spanwise directions (see supplementary movie ``Fig6c\dots mp4"). 
These small fluctuations about the steady single-hump shape somewhat resemble the ``breathing mode" oscillations of membrane disks with edges that are perpendicular to rather than aligned with the oncoming flow~\cite{mathai2023shape}. 

In (d)--(f) the membrane pretension is equal to $10^{-0.75}$, the smallest value of $T_0$ shown in figure~\ref{fig:FFFFar1nu0}. The first color plot in (d) is almost uniform in the horizontal time coordinate which implies that the membrane is an almost-steady, downward-deflected single-hump shape. This motion leads to a spanwise-asymmetric side-to-side oscillation in the second to fourth plots, shown by the zig-zag pattern of the peak value of $\langle z\rangle_c(\alpha_2, t)$. The snapshots for (d) at the top show that the side-to-side oscillation along the span (right column) is coordinated with a fore-aft oscillation along the chord (left column).
Like panel (d), the first color plot of panel (e) shows a spanwise symmetric oscillation about a single hump shape. 
The oscillation amplitude decreases in the second color plot, and gradually a side-to-side motion emerges, shown by the zig-zag patterns of the third and fourth color plots.

Panel (f) shows a similarly complex evolution that is somewhat spanwise asymmetric from the beginning of the first color plot and the asymmetry increases somewhat through the end of the simulation at $t\approx 1000$.

The spanwise-asymmetric motions in the fourth color plots of panels (d), (e), and (f) resemble waves that travel a short spanwise distance before reversing direction, while the symmetric motions in (c) resemble standing waves. Later we will see differences with free side-edges that indicate wave reflection from the fixed side-edges may play a role in these patterns.

\begin{figure}[H]
    \centering
    \includegraphics[width=.91\textwidth]{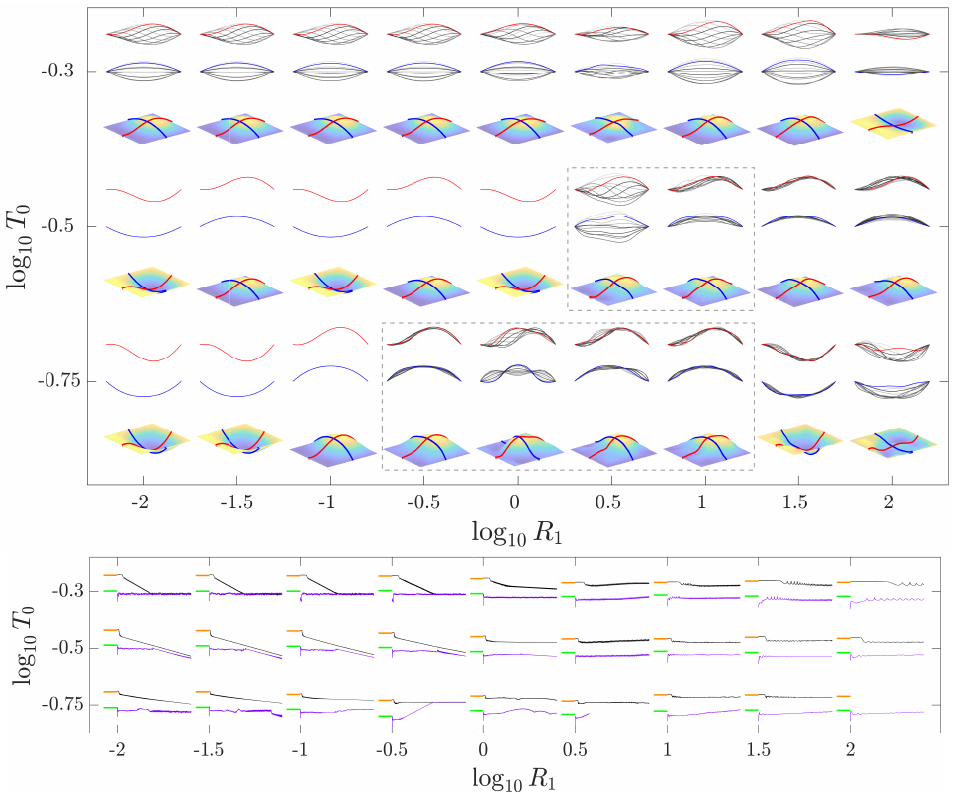}
    \caption{(FFFF) Same quantities as in figure~\ref{fig:FFFFar1nu0} but with $\nu=0.5$.}
    \label{fig:FFFFar1nu05}
\end{figure}

Next, we consider the same FFFF aspect-ratio-one membranes, but increase the Poisson ratio $\nu$ from 0 to 0.5. The Poisson ratio is the ratio of transverse contraction to longitudinal extension during longitudinal stretching. Nonzero values correspond to a volume-conserving property of the material, with $\nu = 0.5$ the upper limit for traditional materials, corresponding to an incompressible solid material, e.g. rubber, approximately~\cite{lakes2001broader}.

In figure~\ref{fig:FFFFar1nu05} we show how the steady-state snapshots differ with Poisson ratio increased to 0.5. 
We now have one up-down symmetric oscillation
at $T_0 = 10^{-0.3}$ that eventually becomes spanwise-asymmetric (at $R_1=10^{0.5}$) and at same $R_1$, one up-down oscillation at 
$T_0 = 10^{-0.5}$, unlike in figure~\ref{fig:FFFFar1nu0}. Also, more of the simulations compute at the smallest $T_0$ value (lower-right corner) with Poisson ratio 0.5.
The types of the dynamics are similar, as is the transition from steady deflections at small $R_1$ to unsteady motions at larger $R_1$, below the $T_0$ for up-down oscillations. One standing-wave motion appears (at $(R_1,T_0 )=(10^{0},10^{-0.75}$)) and short-wavelength modulations across the span are noticeable in more cases with $\nu = 0.5$, particularly at smaller $T_0$. Also, more of the simulations compute successfully at the smallest $T_0$ value (lower-right corner).

At the bottom of figure~\ref{fig:FFFFar1nu05}, the large panel shows many $\log_{10}|z_{\mathrm{asymm}}(t)|$ plots over a large region of $R_1$-$T_0$ space for the asymmetric (black) and symmetric (purple) initial conditions, with the green and orange hash marks denoting  
$\log_{10}|z_{\mathrm{asymm}}| = -7$ and 0 respectively, as in figure~\ref{fig:FFFFar1nu0}. 
At most $(R_1,T_0)$ values, the pattern is similar to the $\nu=0$ FFFF cases in figure~\ref{fig:FFFFar1nu0}. At $T_0\in[10^{-0.5},10^{-0.3}]$ for small-to-moderate values of $R_1$, $|z_{\mathrm{asymm}}(t)|$ decays for both initial conditions. In these cases a symmetric oscillation or steady deflected shape occurs. At $R_1\geq 10^1$, $z_{\mathrm{asymm}}(t)$ generally grows at large times, though the growth rate decreases with increasing $R_1$. The large-time steady-state behavior may only be reached for $t \gg 1000$. 
At $R_1=10^{0.5}$, the $|z_{\mathrm{asymm}}(t)|$ plots are growing slowly at large times for both initial conditions. Moving to $T_0 = 10^{-0.75}$, in contrast to membranes with $\nu=0$ (figure~\ref{fig:FFFFar1nu0}), when $R_1\leq 10^{-1.5}$, the $z_{\mathrm{asymm}}$ decays for both initial conditions. 

\begin{figure}[H]
    \centering
    \includegraphics[width=.96\textwidth]{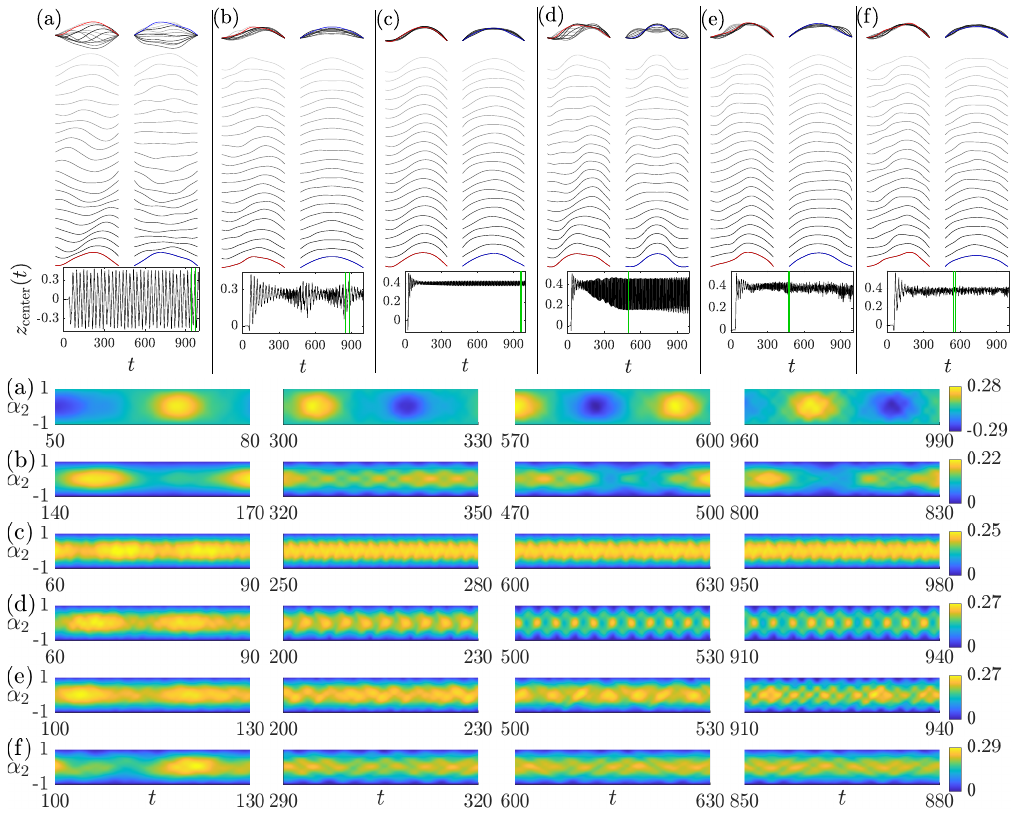}
    \caption{(FFFF) 
    Same quantities as in figure~\ref{fig:ffffar1nu0vertical} but with $\nu=0.5$, and $T_0=10^{-0.5}$ and $R_1$ equal to (a) $10^{0.5}$, (b) $10^{1}$, and $T_0=10^{-0.75}$ and $R_1$ equal to (c) $10^{-0.5}$, (d) $10^{0}$, (e) $10^{0.5}$, and (f) $10^{1}$. The motions in panels (a)--(f) are shown in the supplementary movies named ``Fig8a\dots mp4" through ``Fig8f\dots mp4,'' respectively. 
     }
    \label{fig:ffffar1nu05vertical}
\end{figure}
\FloatBarrier

As with $\nu$ = 0, we study six of the motions in
figure~\ref{fig:FFFFar1nu05} in more detail in 
figure~\ref{fig:ffffar1nu05vertical}, the case in the dashed boxes in
figure~\ref{fig:FFFFar1nu05}. In the first two panels ((a)--(b)) the membrane pretension is equal to $10^{-0.5}$. For (a) only, the membrane undergoes up-down oscillations. The color plots at the bottom of the figure show that in (a), a simple up-down oscillation in the third color plot changes to one with superimposed waves of small wavelength in the fourth color plot that propagate symmetrically across the span. Faint diagonal lines visible in the last color plot indicate the waves, also seen in the midchord snapshots above.
More complicated dynamics are seen in the $z_{\mathrm{center}}(t)$ plot of panel (b), typical of larger $R_1$ values. In the first color plot, a spanwise symmetric oscillation about a single hump shape is seen. The oscillation amplitude decreases and the frequency and spatial complexity increase in the second color plot. Spanwise symmetric motions occur in the third and fourth color plots, with complex temporal patterns. The snapshots for (b) at the top show multiple peaks that move inward and outward along the span (right column), coordinated with peaks that move fore and aft along the chord (left column). Panels (c)--(f) are cases at smaller $T_0$, with intermediate and large $R_1$ that show diverse dynamics. The $z_{\mathrm{center}}(t)$ plot of panel (c) shows a gradual evolution towards a periodic oscillation, and the color plots show a corresponding evolution from a roughly symmetric to a clearly asymmetric side-to-side motion with complex though periodic temporal pattern, as shown in the supplementary movie. 
The membrane in panel (d) at $(R_1,T_0)=(10^0,10^{-0.75})$ has multiple peaks along the midchord (i.e. in the spanwise direction) corresponding to standing waves with fixed nodes and antinodes, synchronized with a fore-aft chordwise oscillation.

At larger $R_1$ ((e) and (f)), the dynamics evolve more slowly, ultimately with spanwise-asymmetric waves. The diagonal bands in the color plots sometimes switch from predominantly positive to negative slope (i.e., the second-to-fourth color plots in (c), (e), and (f)), corresponding to traveling waves that change direction.

\section{Free side-edges}\label{sec:frfr}

We now investigate a different set of boundary conditions: membranes with the leading and trailing edges fixed but the side edges free to move vertically. As in~\cite{mavroyiakoumou2022membrane} we call these FRFR membranes. With the side edges free, the membranes have a wide range of unsteady dynamics, including strongly wavelike motions in the spanwise direction. 

\begin{figure}[ht]
    \centering
    \includegraphics[width=.97\textwidth]{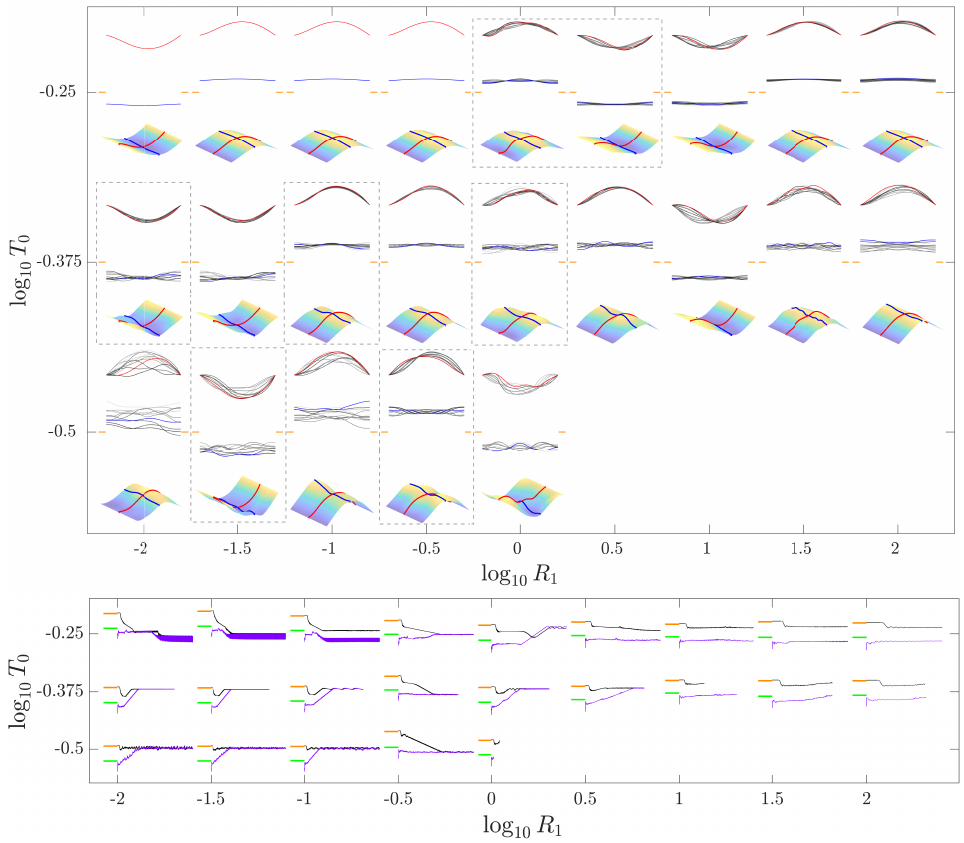}
    \caption{Snapshots of FRFR membranes at large times over a portion of $R_1$-$T_0$ space. The membranes have aspect ratio one and $\nu=0$. At each $(R_1,T_0)$ pair, three sets of snapshots are shown, from top to bottom: midspan snapshots, midchord snapshots, and one 3-D membrane surface snapshot with the midspan and midchord snapshots highlighted in red and blue. The small orange lines next to the midchord snapshots indicate $z=0$. Seven cases, outlined with gray dashed lines, are shown in more detail in figure~\ref{fig:frfrar1nu0vertical}. The large panel at the bottom shows $\log_{10}|z_\mathrm{asymm}(t)|$~\eqref{eq:zasymmetry} over the same portion of $R_1$-$T_0$ space. The plots are given in pairs for the asymmetric (black) and symmetric (purple) initial conditions. At the left end and near the bottom of each pair, the $R_1$ and $T_0$ values are listed on the corresponding axes. The horizontal $t$ and vertical $\log_{10}|z_{\mathrm{asymm}}|$ axes are omitted, but for each pair of plots, the green and orange hash marks denote $\log_{10}|z_{\mathrm{asymm}}|$ at $-7$ and 0, respectively.  }
    \label{fig:FRFRar1nu0}
\end{figure}

The top panel of figure~\ref{fig:FRFRar1nu0} shows sequences of midspan and midchord snapshots, together with a 3-D membrane surface, that represent typical long-time steady-state motions across a range of $R_1$-$T_0$ space. At the lower-right corner ($R_1\geq 10^{0.5}$ with $T_0=10^{-0.5}$), the simulations often fail to converge when the membranes reach large-amplitude deflections so we omit the snapshots there.

We have already seen in \S\ref{sec:ffff} that FFFF membranes can perform spanwise-symmetric up-down oscillations. The top panel in figure~\ref{fig:FRFRar1nu0} shows that up-down oscillations do not occur with free side-edges within the $R_1$-$T_0$ range shown. Instead, membranes oscillate about an upward- or downward-deflected state. Some steady shapes occur at small $R_1$ ($<10^0$) when $T_0=10^{-0.25}$. These single-hump steady motions were also seen in \cite[figure 12]{mavroyiakoumou2022membrane}.
When $T_0$ is decreased to $10^{-0.5}$ and $10^{-0.75}$, spanwise asymmetric motions are typical (seen in the snapshot sequences in the top panel and the O(1) values of $z_{\mathrm{asymm}}$ in the bottom panel), with the only spanwise symmetric cases occurring with $R_1=10^{-0.5}$.
As in the FFFF case, $|z_{\mathrm{asymm}}(t)|$ drops sharply for the asymmetric initial condition when large amplitude is reached. Unlike previously, it then rises in many cases, with both the symmetric and asymmetric initial conditions. For some time periods, particularly at small $R_1$ for both initial conditions, $z_{\mathrm{asymm}}(t)$ increases exponentially, with the late-time dynamics reaching a similar level of asymmetry in the two cases at most $(R_1, T_0)$ values. In contrast, light membranes ($R_1\leq 10^{-1.5}$) with all edges fixed (figure~\ref{fig:FFFFar1nu0}) show spanwise-symmetric oscillations. The constraint of fixed side-edges enforces spanwise symmetry at the side edges, so it is perhaps not surprising that this could make the overall dynamics more spanwise symmetric in some cases. In figure~\ref{fig:FRFRar1nu0}, at $T_0=10^{-0.375}$, the asymmetry growth rate is monotonically decreasing with $R_1$ (with the exception of $R_1=10^{-0.5}$ where the motions are spanwise symmetric). The highest growth rates occur at the smallest $T_0$, $10^{-0.5}$.

\begin{figure}[H]
    \centering
    \includegraphics[width=\textwidth]{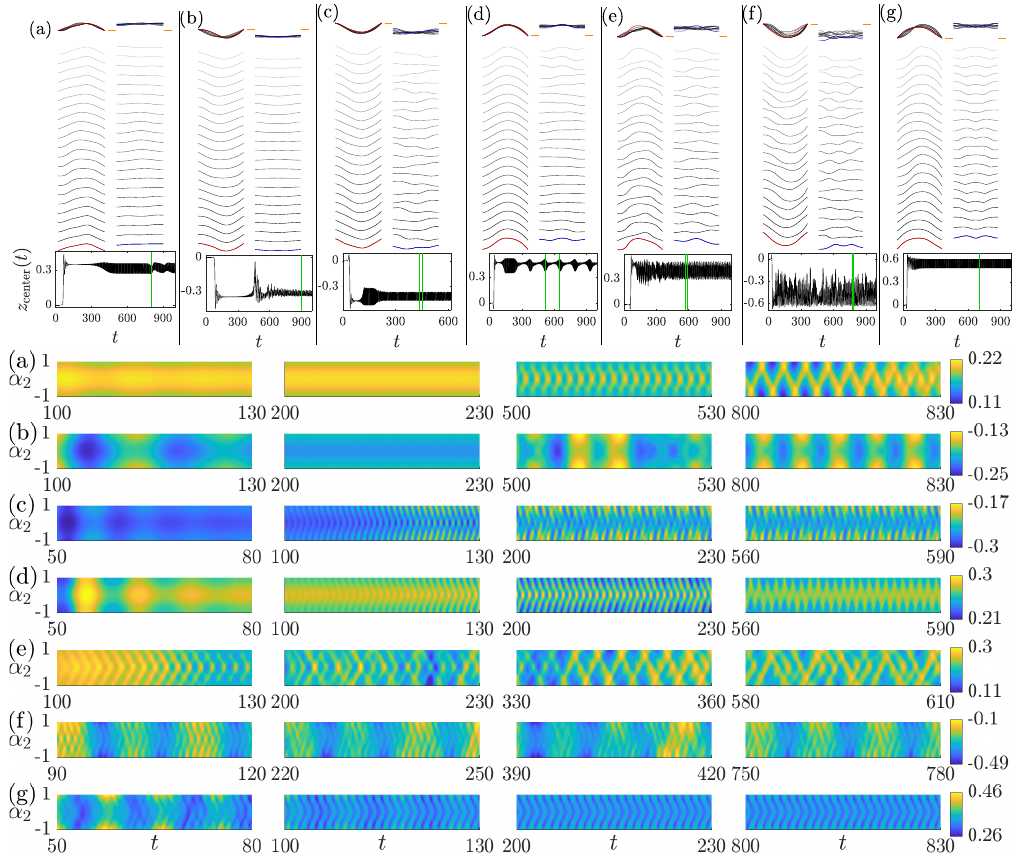}
    \caption{Seven examples of long-time dynamics for FRFR membranes with aspect ratio one and $\nu=0$. Panels (a)--(c) show sequences of midspan (left column) and midchord (right column) snapshots, overlaid at the top of each column, and arrayed vertically below, from top to bottom with increasing time, in darkening shades of gray followed by red and blue at the last times. The snapshots are taken between two relatively late times when the membrane center deflection $z_{\mathrm{center}}$ reaches successive upward or downward extrema. The $(R_1,T_0)$ values are (a) $(10^{0},10^{-0.25})$, (b) $(10^{0.5},10^{-0.25})$, (c) $(10^{-2},10^{-0.375})$, (d) $(10^{-1},10^{-0.375})$, (e) $(10^{0},10^{-0.375})$, (f) $(10^{-1.5},10^{-0.5})$, and (g) $(10^{-0.5},10^{-0.5})$. Below the snapshots, $z_{\mathrm{center}}(t)$ is plotted, with the two green vertical lines marking the time range of the snapshots. At the bottom of the figure the color plots show $\langle z \rangle_c$, the chordwise average of $z$, over selected time intervals labeled at the bottom. The motions in panels (a)--(g) are shown in the supplementary movies named ``Fig10a\dots mp4" through ``Fig10g\dots mp4,'' respectively.}
    \label{fig:frfrar1nu0vertical}
\end{figure}

 Figure~\ref{fig:frfrar1nu0vertical} presents seven of the cases from 
figure~\ref{fig:FRFRar1nu0} (those outlined with gray dashed lines) in more detail. They have all three values of $T_0$---($10^{-0.5}$, $10^{-0.375}$, $10^{-0.25}$)---and six values of $R_1\in[10^{-2},10^{0.5}]$, and represent a variety of different types of dynamics in figure~\ref{fig:FRFRar1nu0}. In panel (a), the motion is an upward-deflected state. The associated $\langle z\rangle_c(\alpha_2,t)$ color plot sequence (a) shows a spanwise-symmetric motion in the form of a single-hump shape when $t\in[100,230]$. The third color plot with $t\in[500,530]$ and the second part of the supplementary movie ``Fig10a\dots mp4" show that the membrane is spanwise symmetric but switches between a shape with two peaks and a trough and a shape with two troughs and a peak along the span. The peaks and troughs move toward and away from the midspan symmetrically on either side of it. In the fourth color plot, the diagonal bands of color correspond to a side-to-side, spanwise-asymmetric motion. In panel (b) the membrane is spanwise symmetric at large amplitudes for the entire simulation. The almost uniform color in the second color plot shows that the membrane is a nearly steady, downward-deflected, single-hump shape that switches to an unsteady motion in the third color plot. A similar motion is observed in the fourth color plot but there the oscillations become nearly periodic in time. Moving to a smaller $T_0$ $(10^{-0.375})$, in panel (c) the membrane first oscillates about the single-hump shape, but then in the second color plot assumes a motion akin to that in the third color plot in (a), an oscillation between spanwise-symmetric shapes with peaks and troughs along the span. Then in the third and fourth color plots, the membrane becomes clearly spanwise asymmetric with diagonal bands that do not extend to the span side-edges. Panel (d) shows the same $T_0$ but larger $R_1$ $(10^{-1})$. The second and third color plots show a spanwise-symmetric oscillation like that in the second color plot of (c), but with an upward-deflected shape instead. The color contrast increases from the second to the third color plot of (d), showing an increasing amplitude of oscillation of the peaks along the span. Shortly thereafter, a spanwise asymmetric oscillation develops (e.g. the fourth color plot), and its amplitude oscillates on a long time scale, $\approx 150$ time units.

In panel (e) the membrane mass is further increased to $10^0$. 
As with panels (c) and (d) for the same $T_0$ value, the membrane goes through multiple spanwise-symmetric states before switching to a spanwise-asymmetric motion with broken bands of color in the third and fourth color plots, showing irregular traveling waves. The oscillation frequency decreases greatly from panels (c)/(d) to (e) as $R_1$ increases from $\ll 1$ to 1.
Moving to the smallest $T_0$, panel (f) shows an  oscillation with larger $z$-amplitude, but still confined to one side (negative $z$).
The first color plot for (f) superposes narrow and wide diagonal bands of color, indicating a small-amplitude high-frequency motion superposed on a larger-amplitude, low-frequency motion. The overall motion is approximately spanwise-symmetric. In the second through fourth color plots, the spanwise asymmetry of the diagonal bands increases. The pattern has regular high-frequency oscillations interspersed with irregular changes over longer time intervals.
The $z_{\mathrm{center}}$ plot above indicates a nonperiodic motion, though the snapshots in the upper panel indicate that ripples move across the span in a somewhat regular pattern. In (g), the first color plot is similar to that in (f) but the second-to-fourth color plots show convergence to different dynamics: a spanwise-symmetric periodic oscillation like those in (c) and (d). The herringbone pattern indicates wave crests and troughs that move inward towards the center of the span, also seen in the snapshots at the top of the figure.

\begin{figure}[H]
    \centering
    \includegraphics[width=\textwidth]{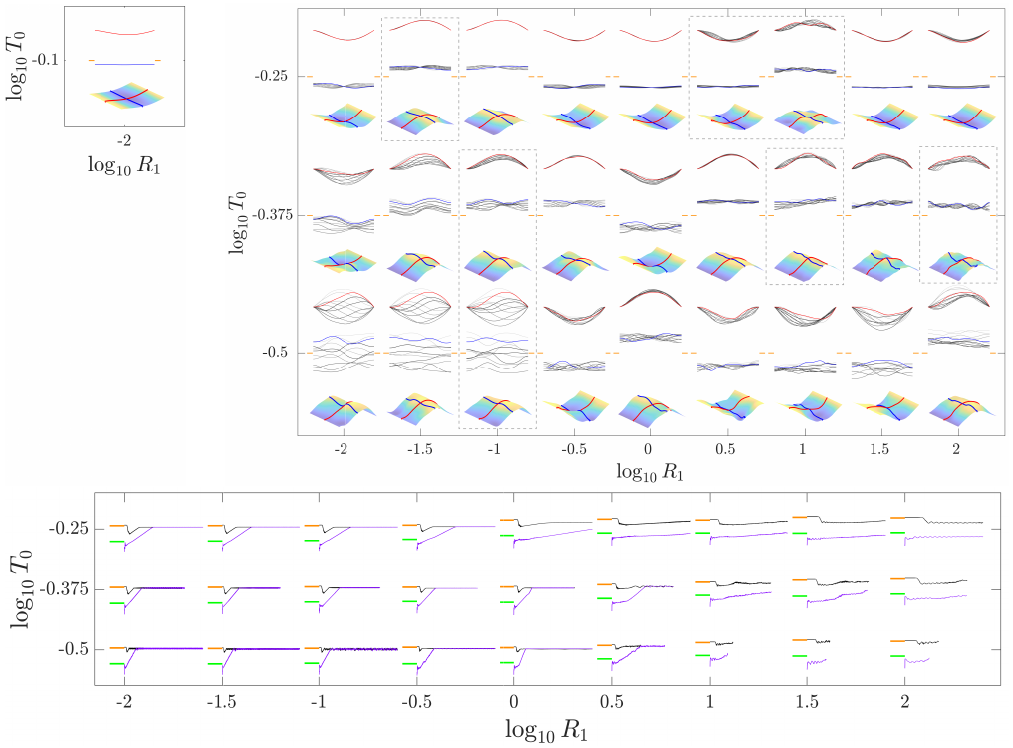}
            \caption{(FRFR) Same quantities as in figure~\ref{fig:FRFRar1nu0} but with $\nu=0.5$. The panel at the upper left with $T_0 = 10^{-0.1}$ is a steady configuration that occurs at all $R_1$ from $10^{-2}$ to $10^2$ at this $T_0$.}
    \label{fig:frfrar1nu05}
\end{figure}

Figure~\ref{fig:frfrar1nu05} shows typical FRFR motions with aspect ratio one but with the Poisson ratio increased to $0.5$. For comparison with $\nu = 0$, the lowest $T_0$ value is set to $10^{-0.5}$, the lowest value at which long-time simulations were obtained across a large range of $R_1$ values in
figure~\ref{fig:FRFRar1nu0}.

At the upper left corner of figure~\ref{fig:frfrar1nu05}, only one panel is shown at $T_0 = 10^{-0.1}$ because the same steady single-hump shape was seen across $R_1$ at this $T_0$. Decreasing $T_0$ to $10^{-0.25}$ and $10^{-0.375}$ (upper right panel), in most cases the membranes instead oscillate about a single-hump shape. At $T_0=10^{-0.25}$, the oscillations are spanwise asymmetric at small $R_1$, with an antisymmetric side-to-side motion in which the midspan is nearly fixed---hence the midspan snapshots are nearly aligned with the red curve. At large $R_1$ and $T_0 = 10^{-0.25}$ and $10^{-0.375}$, the spanwise motions are more complex and sometimes resemble standing waves with multiple fixed nodes or traveling waves. In the third row, $T_0 = 10^{-0.5}$, the membrane performs up-down oscillations at the three smallest $R_1$. At the third smallest $R_1$, $10^{-1}$, the next figure will show that the membrane switches between mainly upward and downward oscillations over $\approx 10$--$100$ time units. At larger $R_1$ the membrane remains in mainly upward and downward oscillations up to $t \approx 1000$. These oscillations are periodic at $R_1 = 10^{-0.5}$ and $10^{0}$ and nonperiodic at larger $R_1$. 
At the largest $R_1$ studied ($R_1\geq 10^1$) and at the smaller values of $T_0$ ($\leq 10^{-0.375}$), the membrane oscillates less regularly, with a shape that is more strongly sloped along the span and more sharply curved along the span and chord.

The bottom panel of figure~\ref{fig:frfrar1nu05} shows $\log_{10}|z_{\mathrm{asymm}}(t)|$ for the spanwise-asymmetric initial condition (black curves), which are the cases shown in the top right panel, and for the spanwise-symmetric initial condition (purple curves). The two initial conditions show similar behavior. 
The bottom panel shows that at most ($R_1$, $T_0$) values, the pattern is similar to the $\nu = 0$ FRFR cases in figure~\ref{fig:FRFRar1nu0}, in which $|z_{\mathrm{asymm}}(t)|$ grows for both initial conditions, but after an early decrease for the asymmetric initial condition when it first reaches large amplitude as a single-hump shape, like in figure~\ref{fig:smallTransientLargeFRFR}(c), before changing to an asymmetric oscillation. The only difference with figure~\ref{fig:FRFRar1nu0} is a few cases ($T_0=10^{-0.25}$ and $R_1\leq 10^{-1}$) where
asymmetric oscillations occur in figure~\ref{fig:frfrar1nu05} rather than the
steady symmetric single-hump shape of figure~\ref{fig:FRFRar1nu0}. The steady symmetric single-hump shape instead
occurs at $T_0=10^{-0.1}$ in figure~\ref{fig:frfrar1nu05}.
By contrast, with FFFF boundary conditions both initial conditions led to spanwise-symmetric oscillations at large times in large-$R_1$ cases ($R_1 \geq 10^{1})$, for both $\nu = 0$ and 0.5.

\begin{figure}[t]
    \centering
    \includegraphics[width=\textwidth]{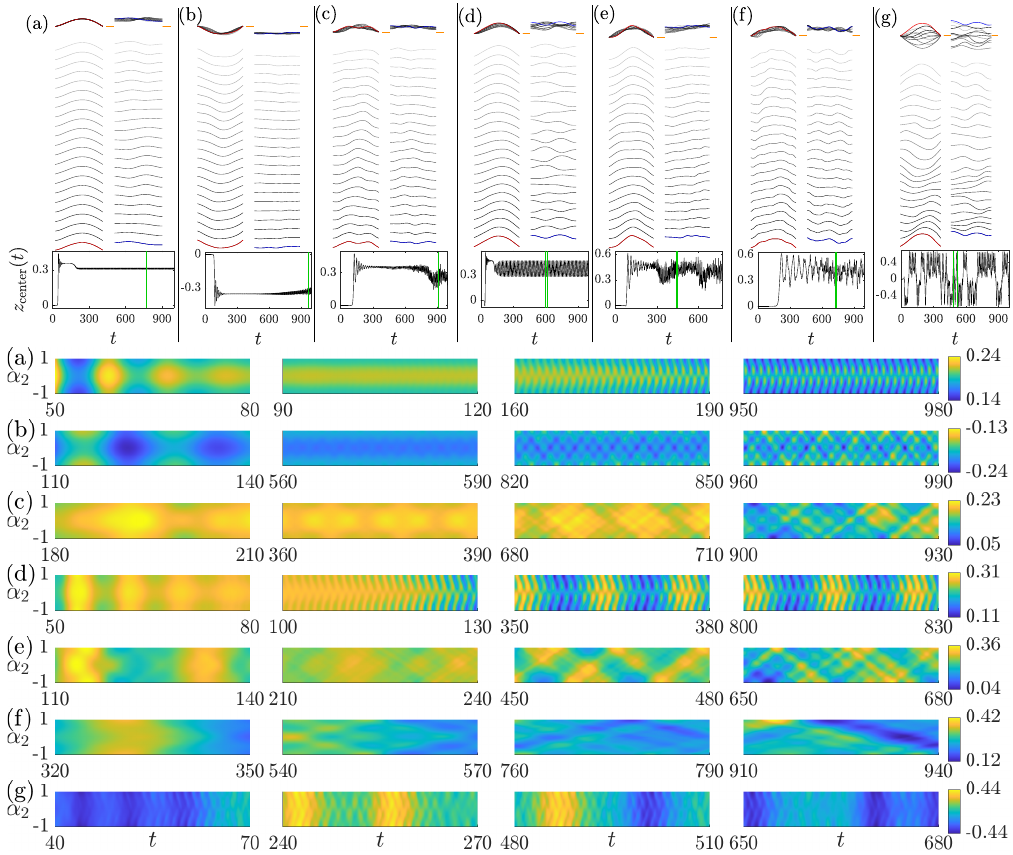}
\caption{(FRFR) Same quantities as in figure~\ref{fig:frfrar1nu0vertical} but with $\nu=0.5$, and $(R_1,T_0)$ values equal to (a) $(10^{-1.5},10^{-0.25})$, (b) $(10^{0.5},10^{-0.25})$, (c) $(10^{1},10^{-0.25})$, (d) $(10^{-1},10^{-0.375})$, (e) $(10^{1},10^{-0.375})$, (f) $(10^{2},10^{-0.375})$, and (g) $(10^{-1},10^{-0.5})$. The motions in panels (a)--(g) are shown in the supplementary movies named ``Fig12a\dots mp4" through ``Fig12g\dots mp4,'' respectively. }
    \label{fig:frfrar1nu05vertical}
\end{figure}

In figure~\ref{fig:frfrar1nu05vertical}, as with $\nu=0$, we study seven of the motions in
figure~\ref{fig:frfrar1nu05} (in the dashed boxes) in more detail. 
In panels (a)--(c) the pretension is equal to $10^{-0.25}$. Panel (a) with $R_1=10^{-1.5}$ shows a motion that evolves from a steady single-hump shape to a periodic side-to-side motion. Panel (b), with much larger mass, enters a similar steady single hump shape initially. This state slowly destabilizes around $t \approx 560$ and leads to a spanwise-asymmetric motion with wavelike features, shown by the diagonal bands in the third and fourth color plots. The motion is close to a periodic oscillation but with nonperiodic features and a growing amplitude at $t = 1000$. With still larger mass, panel (c) has a similar time evolution to panel (b), but the color plots show relatively larger nonperiodic features and a slower wave speed shown by the slopes of the diagonal bands in the color plots.
Moving to smaller $T_0$ in panel (d), the membrane evolves relatively quickly from a symmetric single-hump shape to a periodic asymmetric motion similar to those described previously with $\nu = 0$. The diagonal bands in the color plots show a high-frequency side-to-side oscillation in which extrema on either side of the midspan that move toward it, alternately on one side and then the other. The color plots also show a lower-frequency component in which the overall deflection increases and decreases with time (also shown in the corresponding movie). 
Moving to larger mass in panels (e) and (f), the initial interval of growth from the flat state is longer and the eventual large-amplitude motions are more chaotic, both in 2-D~\cite{mavroyiakoumou2020large} and 3-D~\cite{mavroyiakoumou2022membrane} 
flows. In the first color plots of (e) and (f), the membrane enters the large-amplitude regime having assumed a quasi-periodic state with a membrane deflection that is almost uniform along the span. In the second color plots, the motions are approximately spanwise-symmetric standing waves. 
Then the membrane surface becomes sharper, with more strongly localized traveling waves, shown by the diagonal bands that become more evident in the third and fourth color plots. These bands seem to partially reflect off the free side-edges, with an increase or decrease in the deflection magnitude ($|\langle z\rangle_c(\alpha_2,t)|$) when reflection occurs. In general, the frequency of oscillations decreases from (d) to (e) and from (e) to (f), and the waves become more intermittent, as the membrane mass increases. The decreased frequency corresponds to a wider horizontal spacing between repeated features in the color plots, or a smaller slope of the zig-zag patterns.

Panel (g) shows a motion at the smallest $T_0$ and small $R_1$. In this region of parameter space the membranes undergo up-down oscillations with a side-to-side asymmetric spanwise oscillation.
The membrane switches between negative and positive $z$-values with small oscillations in both the chordwise and spanwise directions at irregular time intervals. The motion is like panel (d) but switches between upward and downward states at irregular time intervals. The smaller oscillations about the upward and downward states are also slightly nonperiodic. The alternating diagonal bands in panels (d) and (g) correspond to side-to-side motions that resemble those seen in the FFFF case, which usually had a mainly upward or downward deflection (e.g.\ figures~\ref{fig:ffffar1nu0vertical}(e) and~(f)).

\FloatBarrier

\section{Effect of aspect ratio}\label{sec:aspectRatio}

Membrane motions may depend strongly on the geometric aspect ratio, i.e. the ratio of the span to the chord length. In nature for example, the aspect ratio plays a critical role in the flight performance of bats. Their wing morphology allows them to switch between large and small aspect ratios depending on their needs: they assume larger aspect ratios for open air and long distance flight, and smaller aspect ratios for slow flight in cluttered environments~\cite{canals2011biomechanical,wang2015lift}.

In~\cite{bleischwitz2015aspect} they describe how a membrane wing's aspect ratio affects aerodynamic performance and membrane deformations for three different low aspect ratios (1, 1.5, and 2). Mean deflection results indicate that lower-aspect-ratio membrane wings show defined U-shape deflections along the span, whereas higher aspect ratios display a progressive rise in deformation to the wing tip.

Studies of 3-D flexible plates indicate that more bodies are unstable at large aspect ratio ~\cite{eloy2007flutter,eloy2008aic,banerjee2015three}, and we found the same for 3-D FRFR and FRRR membranes previously \cite{mavroyiakoumou2022membrane}; as the aspect ratio increases, the stability boundary approaches that of 2-D membranes~\cite{mavroyiakoumou2021eigenmode,mavroyiakoumou2020large}.

In this section, we study the effect of aspect ratio on the membrane dynamics for both FFFF and FRFR boundary conditions. In appendix~\ref{app:StabilityBoundaries} we show the stability boundaries and classification of instability types for the FFFF and FRFR boundary conditions with Poisson ratios 0 and 0.5, and aspect ratios one, two, and four. We now describe how the dynamics compare across the aspect ratios.  In general, as we increase the aspect ratio, more steady single-hump cases occur. For $R_1 \leq 10^{-0.5}$ and aspect ratio 2, nearly all membranes are steady, except for a few cases at the smallest $T_0$, $10^{-0.5}$. For aspect ratio 4, all such cases were steady, though at many parameters the simulations did not run successfully for long times. For $R_1 \geq 1$ and
aspect ratio~2, stable oscillatory motions were found, and these usually had much smaller oscillation amplitudes than for aspect ratio~1. For aspect ratio 4, such cases did not compute successfully, giving unphysically large membrane deflections and curvatures. Due to computational limitations, we used the same number of panels in the spanwise direction for all aspect ratios. It is possible that increasing the number of panels in the spanwise direction would give more successful computations at aspect ratio 4. 

Instead of showing membrane snapshots across $R_1$-$T_0$ space, we give a number of examples of unsteady aspect-ratio-two motions, which usually occur at $R_1 \geq 10^{-0.5}$. We compare these with the corresponding aspect-ratio-one motions in the following four figures, one for each combination of the two boundary conditions and two Poisson ratios.

\begin{figure}[H]
    \centering
    \includegraphics[width=.95\textwidth]{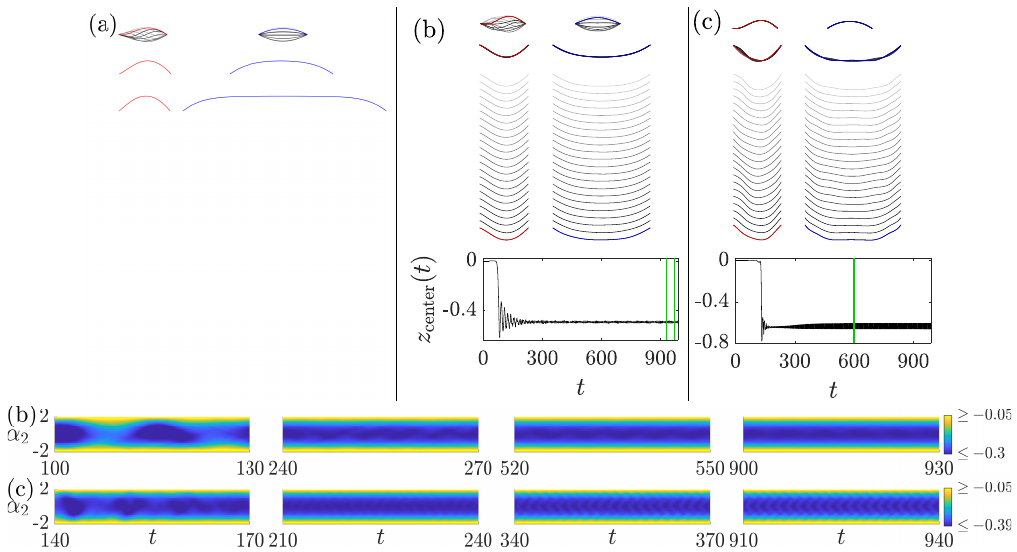}
    \caption{Three examples of long-time dynamics for FFFF membranes with aspect ratio two and $\nu=0$. Panel (a) has $(R_1,T_0)=(10^0,10^{-0.3})$, (b) $(10^1,10^{-0.3})$ and (c) $(10^0,10^{-0.5})$. At the top of each panel are the midspan (left) and midchord (right) aspect-ratio-one snapshots, overlaid. Below are the corresponding aspect-ratio-two snapshots. In~(a) the bottom set of snapshots is for a membrane with aspect ratio four. Farther down the snapshots are vertically arrayed with increasing time moving downward. The aspect-ratio-two (and four) midchord snapshots are dilated by a factor of~2 vertically. Below the snapshots, $z_{\mathrm{center}}(t)$ is plotted, with the two green vertical lines marking the time range of the snapshots. At the bottom of the figure the color plots show $\langle z\rangle_c$, the chordwise average of $z$, over selected time intervals labeled at the bottom. The motions in panels (b) and (c) are shown in the supplementary movies named ``Fig13b\dots mp4" and ``Fig13c\dots mp4,'' respectively.}
\label{fig:verticallyArrayedSnapshotsFFFFsaddleAR4nu0}
\end{figure}

In figure~\ref{fig:verticallyArrayedSnapshotsFFFFsaddleAR4nu0} we show membranes with all edges fixed and Poisson ratio zero. 
We focus on two values of $T_0$, $10^{-0.3}$ and $10^{-0.5}$. In each panel, as in the corresponding figures for aspect-ratio one, we present midspan snapshots (left columns of each panel) and midchord snapshots (right columns of each panel) as well as $z_{\mathrm{center}}(t)$ and $\langle z\rangle_c(\alpha_2,t)$ for the unsteady cases.

Panel (a) compares membrane motions at $R_1=10^{0}$ and $T_0=10^{-0.3}$, for aspect ratios 1, 2, and 4, from top to bottom. In this case and all others, we stretch the $z$ coordinates of the aspect-ratio-two and aspect-ratio-four membranes by a factor of~2 to enhance visibility. 
At these parameters, the aspect-ratio-one membrane motion is a spanwise-symmetric up-down oscillation, whereas the aspect-ratio-two and four membranes have steady, single-hump shapes. The membrane shapes are flatter across the middle part of the span as the aspect ratio increases. In the chordwise direction, the location of the maximum deflection point is slightly towards the trailing edge at aspect ratio 2, and closer to the midchord at aspect ratio 4 (shown by the red snapshots).

In panels (b) and (c), we compare only aspect ratios one (first row of snapshots) and two (second row of snapshots, also vertically arrayed below).
The aspect-ratio-one cases were shown previously in figure~\ref{fig:ffffar1nu0vertical}.
In panel~(b), the dimensionless membrane mass is $R_1=10^1$, larger than (a). Here the aspect-ratio-two membrane is an almost steady shape but with very small oscillations that are periodic in time; seen in both the $z_{\mathrm{center}}(t)$ plot below the vertically arrayed membrane snapshots and in the second-to-fourth color plots at the bottom of the figure. Panel (c) is a case with smaller pretension $(10^{-0.5})$ and here the aspect-ratio-two motion is almost a standing wave with three nodes, whereas the aspect-ratio-one membrane is steady.

\begin{figure}[t]
    \centering
    \includegraphics[width=\textwidth]{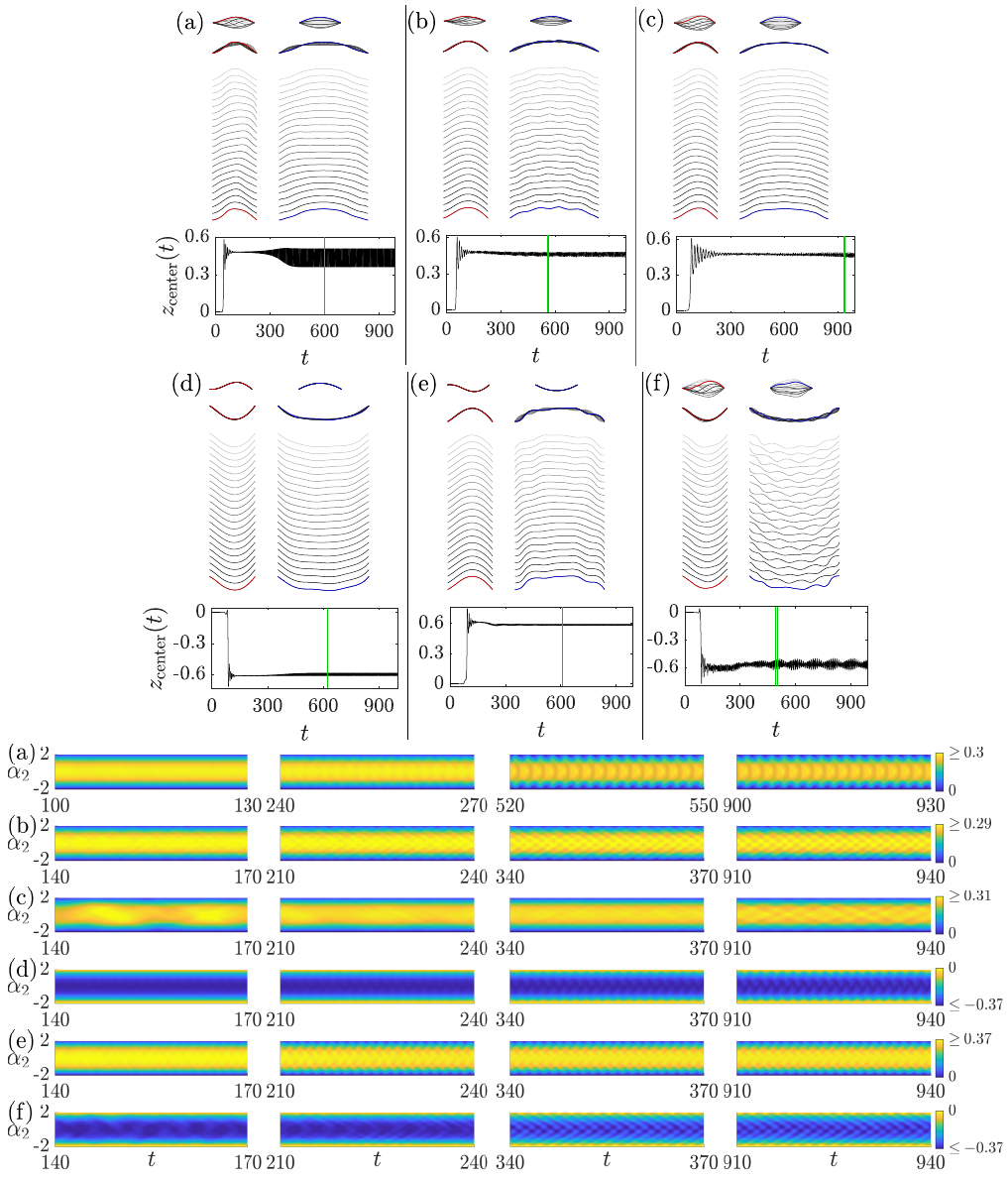}
    \caption{(FFFF) Same quantities as in figure~\ref{fig:verticallyArrayedSnapshotsFFFFsaddleAR4nu0} but for $\nu=0.5$. Panels (a)--(c) have $T_0=10^{-0.3}$ and $R_1$ equal to $10^{0}$, $10^{0.5}$, $10^1$, respectively. Panels (d)--(f) have $T_0=10^{-0.5}$ and $R_1=10^{-0.5}$, $10^0$, and $10^{0.5}$, respectively. The motions in panels (a)--(f) are shown in the supplementary movies named ``Fig14a\dots mp4" through ``Fig14f\dots mp4,'' respectively.}
\label{fig:verticallyArrayedSnapshotsFFFFsaddleAR4nu05}
\end{figure}

We now show unsteady motions of FFFF aspect-ratio-two membranes with $\nu=0.5$ (figure~\ref{fig:verticallyArrayedSnapshotsFFFFsaddleAR4nu05}). Generally there are more unsteady cases compared to the $\nu=0$ case, the previous figure. The only successfully-computed FFFF aspect-ratio-four membranes with $\nu=0.5$ were steady, like panel (a) of the previous figure, and are omitted.

The differences are more striking here than for the $\nu=0$ case. Increasing the aspect ratio from one to two causes membranes to become upward- or downward-deflected shapes with small oscillations, rather than up-down oscillations (when $T_0=10^{-0.3}$ for any $R_1$ (panels (a)--(c)) and when $T_0=10^{-0.5}$ with $R_1=10^{0.5}$ (panel~(f)). The $z_{\mathrm{center}}(t)$ plots show deflections of about 0.5, about 10--30\% greater than those of the corresponding aspect-ratio-one membranes (shown in the $z_{\mathrm{center}}(t)$ plots in figure~\ref{fig:ffffar1nu0vertical}). Relative to the mean of $z_{\mathrm{center}}(t)$, the amplitude of oscillation is generally smaller at aspect ratio two. The color plots in figure~\ref{fig:ffffar1nu0vertical} for aspect ratio one showed a mixture of spanwise symmetric and asymmetric (side-to-side) oscillations, time-periodic but with complex temporal patterns. For aspect ratio two, the color plots in figure~\ref{fig:verticallyArrayedSnapshotsFFFFsaddleAR4nu05} are also a mixture of spanwise symmetric and asymmetric oscillations, with more undulations along the span than at aspect ratio one.

In panel (a), the $z_{\mathrm{center}}(t)$ plot shows that the membrane first becomes approximately steady at large amplitude before switching to periodic spanwise-symmetric oscillations about the single-hump shape, shown by the vertically-arrayed midchord snapshots in the right column of panel (a) and the third and fourth color plots of $\langle z\rangle_c(\alpha_2,t)$ below. The membrane alternates between a flat shape and one with the peak of deflection sharply concentrated toward the midspan.
The vertically-arrayed midspan snapshots in the left column of panel (a) and the accompanying movie (``Fig14a\dots mp4") show that the membrane's maximum deflection point is closer to the membrane's trailing edge when the membrane is flatter.

The unsteadiness along the midspan profiles diminishes as the mass is increased to $10^{0.5}$ in (b) and to $10^{1}$ in (c) for the same pretension $T_0$. Meanwhile the number of peaks and troughs along the span increases.    
The snapshots in panels (d) and (e), with smaller pretension ($10^{-0.5}$), show  membrane dynamics that are spanwise asymmetric, while the corresponding aspect-ratio-one membranes are steady.
The midchord snapshots in panel~(f), with larger $R_1$, show a larger-amplitude motion with more undulations along the span.
The corresponding aspect-ratio-one motion is an up-down oscillation which also has complex spatial variations along the span.
The color plots in (b), (d) and (e)  show spanwise asymmetry that is not present for the same parameters when the aspect ratio is one.

\FloatBarrier

\begin{figure}
    \centering
    \includegraphics[width=\textwidth]{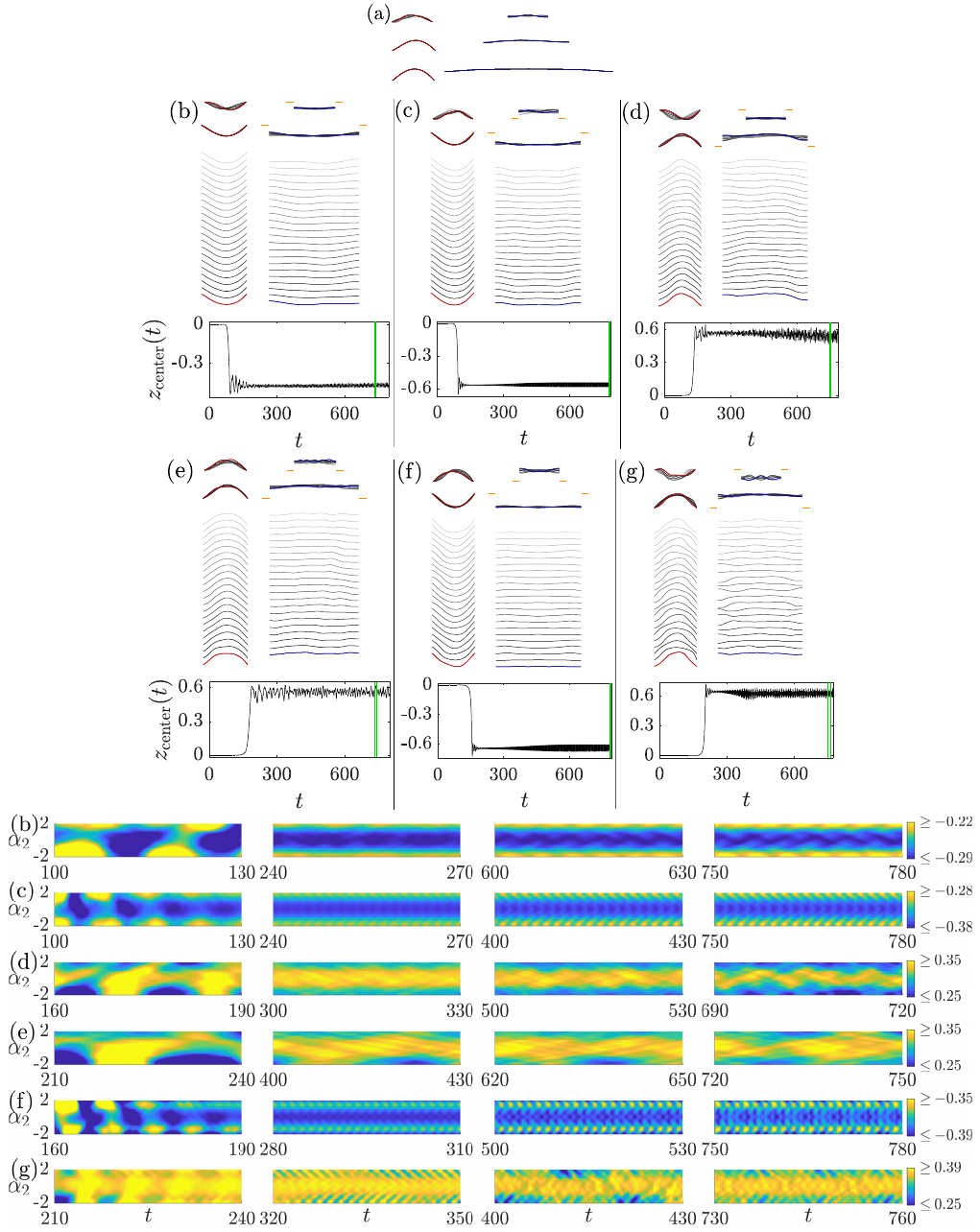}
    \caption{ 
Same quantities as in figure~\ref{fig:verticallyArrayedSnapshotsFFFFsaddleAR4nu0} but for FRFR aspect-ratio-two membranes with $\nu=0$ (compared with snapshots of aspect ratio one and four membranes). In (a)--(b), $T_0=10^{-0.25}$ and $R_1$ is equal to $10^0$ and $10^{1}$, respectively. In (c)--(e), $T_0=10^{-0.375}$ and $R_1$ is equal to $10^{0}$, $10^{1}$, and $10^{1.5}$, and in panels (f)--(g), $T_0=10^{-0.5}$ and $R_1$ is equal to $10^{-0.5}$ and $10^0$, respectively. The motions in panels (b)--(g) are shown in the supplementary movies named ``Fig15b\dots mp4" through ``Fig15g\dots mp4,'' respectively. }
    \label{fig:verticallyArrayedSnapshotsFRFR}
\end{figure}

We now discuss the dynamics with large aspect ratio for free side-edges (FRFR membranes). In~\cite{mavroyiakoumou2022membrane} we found 
that the shapes of 1-D curvilinear fixed--fixed membranes in 2-D flow agreed well with those of FRFR membranes at large aspect ratio (i.e., 8). However, we used only 10 panels along the span, which was not sufficient to resolve the spanwise dynamics. Now we study them in more detail with 40 panels along the span. We use all three pretension values, $T_0\in\{10^{-0.5},10^{-0.375},10^{-0.25}\}$, to compare with the aspect-ratio-one results in figure~\ref{fig:FRFRar1nu0}. In figure~\ref{fig:verticallyArrayedSnapshotsFRFR} we show membrane dynamics for four $R_1$ values ranging from $10^{-0.5}$ to $10^{1.5}$. As in the previous figure, we compare overlaid snapshots of aspect-ratio-one membranes (top row) with aspect-ratio-two membranes (second row) in each panel. Panel (a) compares membranes with aspect ratios one, two, and four, and as in the FFFF case, the profiles are flatter along the span with larger aspect ratio (blue curves), while the profiles become more fore-aft symmetric, shown by the red curves. 
The vertically-arrayed midchord snapshots and the color plots of $\langle z\rangle_c$ below show spanwise-symmetric oscillations in panels (b) and (f), and asymmetric oscillations in the other panels. The oscillation amplitudes are usually larger at aspect ratio one, as seen by comparing the midspan snapshots (red).

With large $R_1$ ($10^1$ in (d) and $10^{1.5}$ in (e)) the membranes become chaotic and spanwise asymmetric, with fluctuations that are similar in magnitude with the corresponding cases with aspect ratio one. The color plots for (d) and (e) show diagonal bands at irregular times and locations, but with slopes that are fairly consistent across the last three time intervals. The slopes of diagonal bands decrease consistently from (d) to (e), consistent with a decrease in the speeds of wavelike features. In panel (g), the standing wave motion of the aspect-ratio-one membrane becomes an asymmetric motion with smaller oscillations at aspect ratio two.  In the FFFF case, the aspect ratio had a clear effect on the fore-aft symmetry of the midspan snapshots. In the FRFR cases the effect is less apparent.

\begin{figure}
    \centering
    \includegraphics[width=\textwidth]{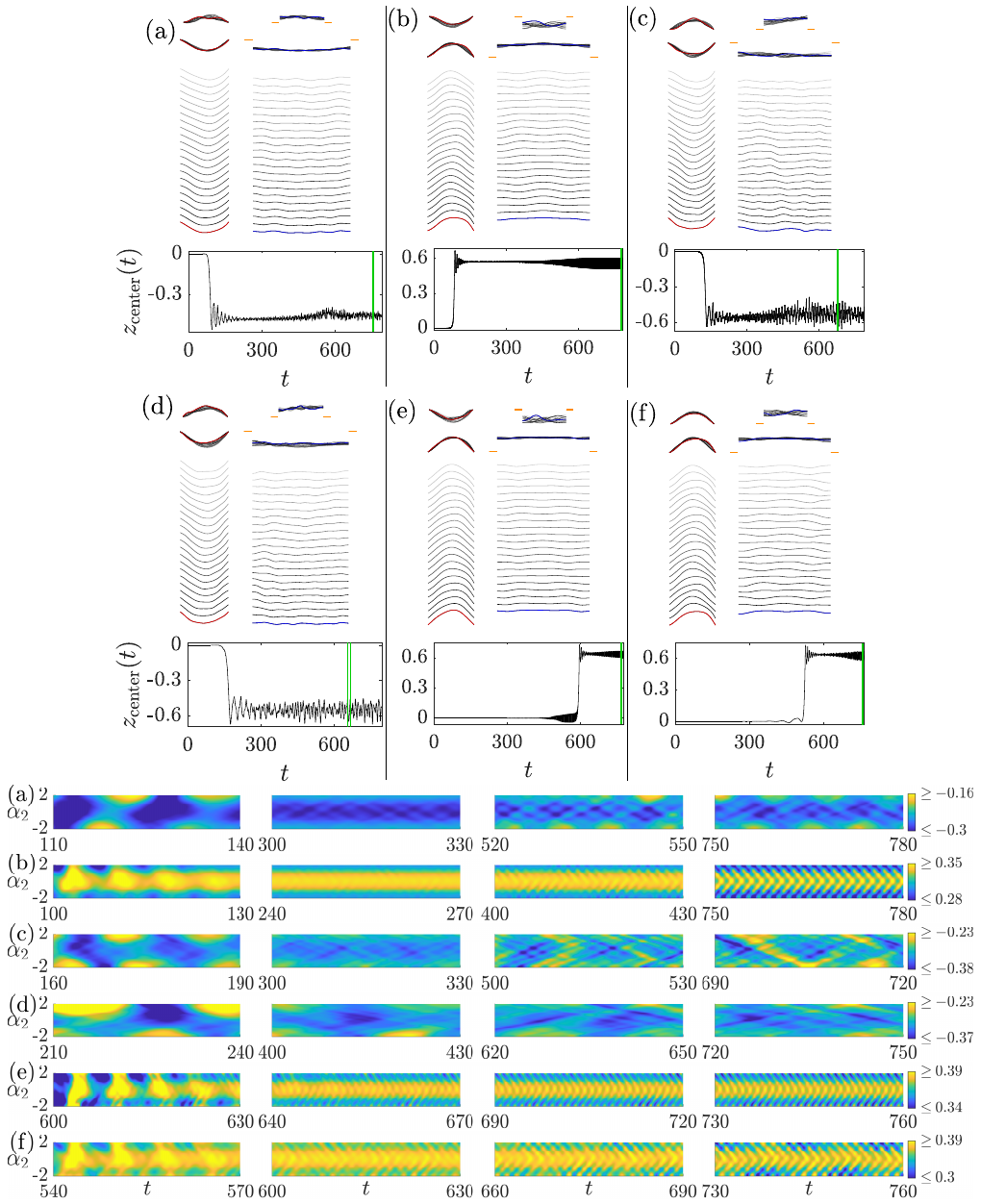}
     \caption{(FRFR) Same quantities as in figure~\ref{fig:verticallyArrayedSnapshotsFRFR}(b)--(g) but with $\nu=0.5$. The motions in panels (a)--(f) are shown in the supplementary movies named ``Fig16a\dots mp4" through ``Fig16f\dots mp4,'' respectively.}
    \label{fig:verticallyArrayedSnapshotsFRFRnu05}
\end{figure}

In figure~\ref{fig:verticallyArrayedSnapshotsFRFRnu05} we study FRFR motions at aspect ratio two with $\nu$ increased to 0.5. 
The aspect-ratio-four membranes are similar to the case shown in figure~\ref{fig:verticallyArrayedSnapshotsFRFR}(a) so we omit them here.
Panels (a)--(f) use the same parameters as in figure~\ref{fig:verticallyArrayedSnapshotsFRFR}(b)--(g): $T_0=10^{-0.25}$ in (a), $10^{-0.375}$ in (b)--(d), and $10^{-0.5}$ in (e)--(f). For aspect ratio two, the color plots in figure~\ref{fig:verticallyArrayedSnapshotsFRFRnu05} are also a mixture of spanwise symmetric and asymmetric oscillations, with some variations in $\langle z\rangle_c(\alpha_2,t)$ along the span (shown by the color bar ranges) that are nonetheless spatially complex, particularly in (c) and (d).
In panel (a) there is little change of the midspan snapshots (left columns) at aspect ratio two.
Panel (b) is almost a standing wave motion with 3 nodes. In (c) and (d) the membranes are less sloped along the span than at aspect-ratio one and have peaks and troughs that move slowly along the span. Similar to figure~\ref{fig:smallTransientLargeFRFR}(c) the motions in panels (e) and (f) first become a two-bump shape in the large-amplitude regime at $t\in[10,550]$---with the peak and trough along the span. These two-bump motions are shown in the initial time intervals of supplementary movies named ``Fig16e\dots mp4" and ``Fig16f\dots mp4,'' respectively. They then resemble standing wave motions (with the same number of nodes as in (b)) but with slightly more heterogeneity in (f) along the span, and with much smaller oscillations compared to the corresponding membranes with aspect ratio one. In (f), the aspect-ratio-one membrane has an antisymmetric oscillation while the aspect-ratio-two membrane performs a spanwise-asymmetric motion. In summary, the FRFR membranes with $\nu = 0.5$ and aspect ratio two usually show smaller oscillations, with more regular spatial structures, than those at aspect ratio one.

\FloatBarrier
\section{Different spanwise and chordwise pretensions} \label{sec:spanAndChordPrestrain}

In the previous sections, we assumed the pretension $T_0$ is isotropic---i.e., it has the same value in all directions. The membrane skin in a bat wing is anisotropic, with prestretched elastin fibers along the spanwise direction that may aid in preventing flutter and improving flight efficiency~\cite{cheney2015wrinkle,lauber2023rapid}. 
Another example occurs in wingsails, where the chordwise and spanwise pretensions are chosen so as to minimize deflections due to aerodynamics forces~\cite{ormiston1971theoretical}. 
A spanwise and chordwise anisotropy in wing compliance may be useful for controlling flutter with a turbulent boundary layer~\cite{duncan1992generation}.
Many types of anisotropy could be considered, but for simplicity, we consider only the case of pretension values $T_{0s}$ and $T_{0c}$ in the spanwise and chordwise directions respectively that are distinct, i.e. $T_{0s}\neq T_{0c}$. 

For comparison with previous results we still fix the stretching rigidity $R_3$ as $1$, and therefore~\eqref{eq:T0cformula} and~\eqref{eq:T0sformula} become:
\begin{align}
    T_{0c} &= \frac{1}{2(1-\nu^2)}(1+\overline{e}_c)^2\left[\left((1+\overline{e}_c)^2-1\right)+\nu \left((1+\overline{e}_s)^2-1\right)\right],\\
    T_{0s}& =  \frac{1}{2(1-\nu^2)}(1+\overline{e}_s)^2\left[\left((1+\overline{e}_s)^2-1\right)+\nu \left((1+\overline{e}_c)^2-1\right)\right].
\end{align}
We first consider membranes with all edges fixed and aspect ratio one (as in~\S\ref{sec:ffff}) and choose $\nu=0$. We consider two very anisotropic cases in which the chordwise pretension is either much larger than or much smaller than the spanwise pretension: $(T_{0c},T_{0s}) = \left(10^{-0.3},10^{-0.75}\right)$ and
$\left(10^{-0.75},10^{-0.3}\right)$, respectively. These pretension values are the extremes of the $T_0$ values in figure~\ref{fig:FFFFar1nu0}. 
\begin{figure}[H]
    \centering
    \includegraphics[width=.85\textwidth]{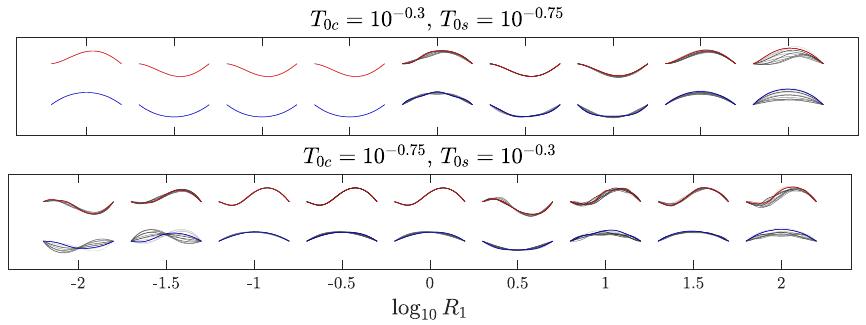}
    \caption{A comparison of FFFF membrane dynamics with anisotropic pretension. Here the aspect ratio is one, $\nu=0$, $T_{0c}=10^{-0.3}$ and $T_{0s}=10^{-0.75}$ (in the top panel) and $T_{0c}=10^{-0.75}$ and $T_{0s}=10^{-0.3}$ (in the bottom panel). Overlaid sequences of midspan (in the first row within each panel) and midchord (in the second row within each panel) snapshots of large-amplitude membrane motions.}
    \label{fig:FFFFar1nu0T0cT0s}
\end{figure}

In figure~\ref{fig:FFFFar1nu0T0cT0s}
we show sequences of midspan and midchord snapshots across the same range of $R_1$ used previously.
In the isotropic case with $T_0=T_{0s}=T_{0c}=10^{-0.3}$ in figure~\ref{fig:FFFFar1nu0}, the membranes performed up-down oscillations with upstream traveling waves in the chordwise direction and were spanwise symmetric. At $T_0=10^{-0.75}$ the membranes assumed a single-hump shape, either steady or with small oscillations that were spanwise asymmetric. 
The spanwise profiles (blue) of the anisotropic cases shown in figure~\ref{fig:FFFFar1nu0T0cT0s} are 
surprisingly similar to each other at moderate to large values of $R_1$ ($\in[10^{0},10^{1.5}]$) where the membranes oscillate about an upward or downward deflection. The chordwise profiles (red) are more clearly different, with a small bump near the leading edge in the bottom row, $T_{0s}>T_{0c}$. The spanwise profiles (blue) mostly differ at small $R_1$ where the membranes are almost steady when $T_{0c}>T_{0s}$,  and instead have periodic side-to-side oscillations when $T_{0c}<T_{0s}$. The motions in figure~\ref{fig:FFFFar1nu0T0cT0s} resemble those with isotropic pretension $T_0$ between $10^{-0.5}$ and $10^{-0.75}$. 

\begin{figure}[H]
    \centering
    \includegraphics[width=.85\textwidth]{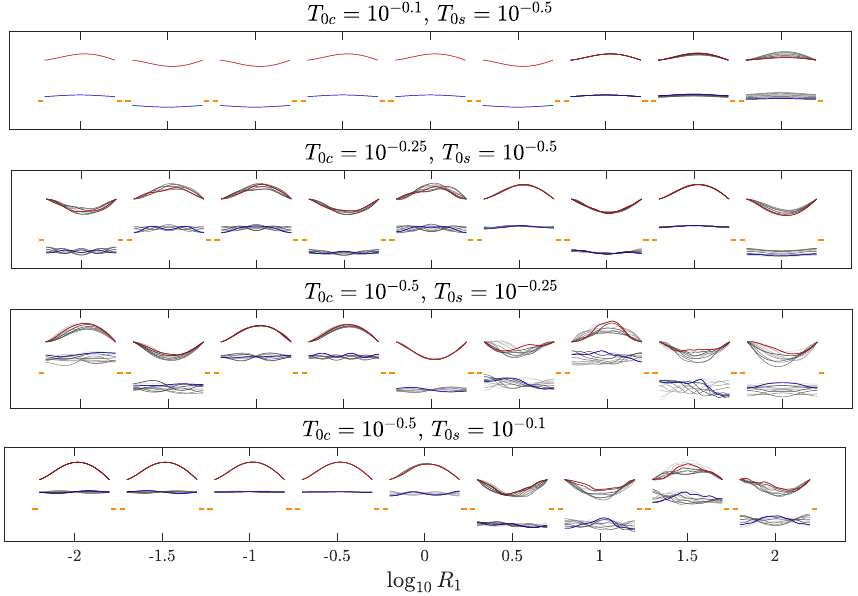}
    \caption{Same as figure~\ref{fig:FFFFar1nu0T0cT0s} but for FRFR membranes with $\nu=0.5$, $T_{0c}=10^{-0.1}$ and $T_{0s}=10^{-0.5}$ (in the top panel), $T_{0c}=10^{-0.25}$ and $T_{0s}=10^{-0.5}$ (in the second panel), $T_{0c}=10^{-0.5}$ and $T_{0s}=10^{-0.25}$ (in the third panel) and $T_{0c}=10^{-0.5}$ and $T_{0s}=10^{-0.1}$ (in the bottom panel).}
    \label{fig:FRFRar1nu05T0cT0s}
\end{figure}
We consider anisotropic pretensions for FRFR membranes in figure~\ref{fig:FRFRar1nu05T0cT0s} and choose $\nu=0.5$ for variety. Here we take different combinations of $(T_{0c},T_{0s})$ in the range $\left(10^{-0.5},10^{-0.25},10^{-0.1}\right)$, three of the $T_0$ values in the corresponding isotropic case, figure~\ref{fig:frfrar1nu05}.
In that figure, with $T_0=T_{0s}=T_{0c}=10^{-0.1}$, the membranes were all steady single-hump shapes with small deflections. At $T_0=10^{-0.5}$ the membranes were all unsteady and spanwise-asymmetric, with up-down oscillations at the three smallest $R_1$ and large-amplitude oscillations about a single-hump shape at most other $R_1$. In the first row of figure~\ref{fig:FRFRar1nu05T0cT0s}, $T_{0c} \gg T_{0s}$, the membranes are steady at small-to-moderate $R_1$ but still in the process of converging to steady states at the latest times at large $R_1$.
The behavior matches that with isotropic pretension equal to the value of $T_{0c}$, $10^{-0.1}$.  
In the second row, $T_{0c}=10^{-0.25}$ and $T_{0s}=10^{-0.5}$, there are unsteady oscillations that are usually spanwise symmetric. Spanwise symmetric oscillations were not seen with isotropic pretension, but the oscillation amplitudes and chordwise profiles resemble those with isotropic pretensions of $10^{-0.25}$ and $10^{-0.375}$. In the third row, the chordwise and spanwise pretensions are reversed. The oscillations are similar to those in the second row but have larger amplitudes and more spanwise asymmetry.
In the fourth row, $T_{0c}=10^{-0.5}$ and $T_{0s}=10^{-0.1}$, the membranes at small $R_1$ perform an antisymmetric side-to-side motion in which the midspan is nearly fixed. At larger $R_1$ the membranes are more unsteady with larger deflection amplitudes. The
motions in this panel resemble isotropic cases with $T_0 = 10^{-0.25}$ (at small $R_1$) and $T_0 = 10^{-0.5}$ (at moderate and large~$R_1$). 

Like the anisotropic FFFF cases of the previous figure, the anisotropic FRFR motions lie within the range of isotropic motions with $T_0$ between the maximum and minimum values of $T_{0s}$ and $T_{0c}$. The anisotropic FRFR motions often resemble the isotropic motions with $T_0 = T_{0c}$, but not always. A more definite conclusion may require results at more combinations of anisotropic pretensions together with $R_1$, but we do not pursue this here. 

\section{Conclusions}\label{sec:conclusions}

We have studied the variety of spanwise dynamics that can occur in membranes that are initially aligned with steady 3-D inviscid fluid flows and become unstable by flutter or divergence instabilities. We focused on rectangular membranes that are fixed at the leading and trailing edges and either fixed or free at the side edges. With fixed side-edges any nonflat membrane configuration is nonuniform along the span. Its shape may be relatively simple, such as the single-hump shape often studied previously, or have more complex and unsteady dynamics. With free side-edges, the dynamics may be uniform along the span, but are often highly nonuniform and unsteady.  
The FFFF and FRFR boundary conditions are relatively simple and do not explicitly impose complicated spanwise dynamics. In fact, the fixed leading and trailing edges inhibit variations in spanwise deflection. Nonetheless, complex spanwise variations often occur, which indicates that they would also occur for many other boundary conditions including the ten in \cite{mavroyiakoumou2022membrane} that we did not consider here.

In \S\ref{sec:initialConditions} we found a surprising variety of dynamics simply by varying the small initial perturbation between two options, one symmetric and one asymmetric about the midspan cross-section. We expected the membranes to quickly assume the shape of the fastest-growing mode (like those in our 2-D eigenmode study \cite{mavroyiakoumou2021eigenmode}) in both cases. Instead the different initial conditions led to transitions from small to large deflections and transient large-amplitude motions that differed over long times (hundreds of time units) before reaching the same steady-state motion. With the asymmetric initial condition, membranes typically remained asymmetric with complex wavy fluctuations at small amplitude. At aspect ratio one, they then typically assumed a symmetric (single-hump) shape when they first reached large amplitude. The shape either remained symmetric or transitioned to a spanwise-asymmetric motion. At aspect ratio four, the membranes instead reached large amplitude with a simple steady asymmetric shape---with one upward bump and one downward bump along the span---which destabilized and then assumed a steady symmetric single-hump shape, often after more than 100 time units. The symmetric initial condition instead remained symmetric throughout the transition to large amplitude.

The FFFF membranes (all edges fixed) typically had up-down oscillations at large pretensions, close to the stability boundary. At smaller pretensions, steady upward or downward shapes or moderate oscillations about those shapes were seen. These oscillations sometimes had a ``side-to-side" asymmetric motion but spanwise-symmetric oscillations also occurred. The motions were usually simple or more complex periodic oscillations at smaller membrane mass, and usually chaotic motions at larger mass. With the Poisson ratio increased from 0 to 0.5, an up-down oscillation was also observed at a smaller pretension and with spanwise asymmetry. Below this pretension, more cases computed successfully at large values of $R_1$ and the pattern of dynamics was generally similar to that seen at Poisson ratio 0, with the additions of a standing wave motion, with nodes and antinodes along the span.

The FRFR membranes (side edges free) with aspect ratio one and Poisson ratio 0 assumed a single hump shape when the pretension was slightly below the stability boundary. These shapes were steady at small $R_1$ and had small spanwise-symmetric oscillations at moderate and large $R_1$. 
At smaller $T_0$, the motions were all unsteady, with larger oscillation amplitudes and more spanwise asymmetry. The oscillations remained one-sided (as opposed to up-down) at the smallest $T_0$, $10^{-0.5}$. As in the FFFF case, side-to-side oscillations along the span were seen. Traveling wave motions were also seen, identified by diagonal bands in color plots of the chordwise average of the $z$-deflection versus the spanwise coordinate and time. 
The oscillation frequencies and traveling wave speeds varied inversely with the membrane mass.

Increasing the Poisson ratio to 0.5 for the FRFR membranes, the membrane shapes and motions were generally more spanwise asymmetric than at Poisson ratio 0. The oscillation amplitudes were larger, and up-down oscillations appeared at the smallest $T_0$, $10^{-0.5}$.
As for FFFF, more cases were computed successfully at the smallest pretension and large membrane mass with Poisson ratio 0.5.

Increasing the aspect ratio to two, the membranes at small $R_1$ became steady but at moderate-to-large values of $R_1$ they remained unsteady, with a mixture of spanwise symmetric and asymmetric shapes, although with flatter spanwise deflection profiles compared to the aspect-ratio-one cases. This tendency towards a more spanwise-uniform deflection at higher aspect ratio agrees with previous simulations of flags of different aspect ratios in 3-D flows \cite{yu2012numerical}. At aspect ratio two the oscillations had smaller amplitudes but often had more spatial complexity in the waves of deflection along the span. The FRFR cases showed traveling wave motions with multiple crests along the span simultaneously, moving in different directions, and colliding or dispersing at various times.
At aspect ratio four, fewer membranes computed successfully. All those that did assumed a steady single-hump shape, symmetric and flatter along the span than at aspect ratios one and two. 

We briefly considered the effect of different pretensions in the chordwise and spanwise directions. By comparing the dynamics with chordwise and spanwise pretensions set to small and large values respectively, and vice versa, we found that the dynamics were often qualitatively different in the two cases. However, the dynamics resembled cases with isotropic pretension, at values in the range between the anisotropic values.

Finally, we return to the question from the introduction that asked when membrane motions are essentially 2-D versus 3-D. Essentially all cases with aspect ratios two and four (broadly in $R_1$-$T_0$ space and beyond the examples shown here) have deflections that are close to spanwise uniformity in the steady-state motion, except near the side edges in the FFFF cases, where the deflection decays smoothly to zero. Interesting small wavelike motions occur at moderate and large $R_1$ that may be important in some applications. At aspect ratio one strong spanwise variations are typical, though many cases are steady single-hump shapes or have small oscillations about
such shapes. In these cases the flows may resemble 2-D flows but modified by a simple spanwise variation of flow quantities (such as fluid pressure) along the span \cite{eloy2007flutter}. Such cases occur more often at small $R_1$, but their distribution with respect to $R_1$ and $T_0$ does not follow a simple general pattern. Another class of motions that often have simple deflection profiles along the spanwise
direction are the up-down oscillations in the FFFF case. Nonetheless, the corresponding flow seems difficult to approximate using quantities from an unsteady 2-D~flow.

\section*{Acknowledgments}
S.A. acknowledges support from the NSF-DMS Applied Mathematics program,
award number DMS-2204900.



\section*{Supplementary material}
Supplementary movies along with a caption list are available at \url{https://drive.google.com/drive/folders/1HBwh7zRdC_Tq-i7AM55qxTqWQjlPGiUn?usp=sharing}.

\appendix
\section{Thin-membrane elasticity}\label{app:moreOnModel}
Here we derive the expression for the elastic stretching forces based on nonlinear shell theory \cite{koiter1966nonlinear,efrati2009elastic}. A linearized approximation of this expression was used in~\cite{mavroyiakoumou2022membrane}.

We compute the stretching energy of the membrane using the position $\mathbf{r}(\alpha_1,\alpha_2,t)$. The membrane has thickness $h\ll L,W$, the lateral dimensions. We assume the stretching strain is constant through the thickness, accurate to leading order in $h$ \cite{efrati2009elastic}.

We denote the flat pre-strained configuration of the membrane central surface by $\bm{\alpha} \equiv (\alpha_1,\alpha_2,0) = \mathbf{r}(\alpha_1,\alpha_2,0)$. A small line of material connecting two material points $\bm{\alpha}$ and $\bm{\tilde{\alpha}}$ in the flat pre-strained configuration is $\d\bm{\alpha} = \bm{\alpha} - \bm{\tilde{\alpha}}$.
We assume that the pre-strained state is obtained from the zero-energy state by applying uniform pre-strains  $\overline{e}_c$ and $\overline{e}_s$ in the chordwise and spanwise directions, respectively.
Thus in the zero-energy state the small line of material is $(\d\alpha_1/(1+\overline{e}_c), \d\alpha_2/(1+\overline{e}_s))$, the pre-strains having been removed by dividing by $(1+\overline{e}_c)$ and $(1+\overline{e}_s)$.

One of the most common measures of deformation in nonlinear elasticity is the difference between the squared length of a material line in the deformed and zero-energy configurations \cite{landau:1986a,tadmor2012continuum}:
\begin{equation}\label{eq:2eps}
\|\d\mathbf{r}\|^2-\|(\d\alpha_1/(1+\overline{e}_c), \d\alpha_2/(1+\overline{e}_s))\|^2=2\epsilon_{ij}\,\d \alpha_i\d \alpha_j\;,\quad \epsilon_{ij}=\frac{1}{2}\left(a_{ij}-\bar{a}_{ij}\right)
\end{equation}
where $\d\mathbf{r} = \mathbf{r}(\alpha_1,\alpha_2,0) - \mathbf{r}(\tilde{\alpha}_1,\tilde{\alpha}_2,0)$ and $\epsilon_{ij}$ is the (Green-Lagrange) strain tensor, written in terms of the metric tensor $a_{ij}$ and its zero-energy state $\bar{a}$
\begin{equation}
a_{ij}=\partial_{\alpha_i}\mathbf{r}\cdot\partial_{\alpha_j}\mathbf{r}=\frac{\partial r_k}{\partial \alpha_i}\frac{\partial r_k}{\partial \alpha_j} \, ,\; i,j=1,2 ; \; 
\bar{a}_{ij} \equiv \left(\begin{array}{cc}
(1+\overline{e}_c)^{-2} & 0  \\
0 & (1+\overline{e}_s)^{-2}  \\
 \end{array}\right).
\label{eq:gtensor}
\end{equation}

For an isotropic membrane with Young's modulus $E$, Poisson ratio $\nu$, and thickness $h$, the elastic energy per unit midsurface area~\cite{alben2019semi} is
\begin{equation}\label{eq:wAepsilon}
    w_s=\frac{h}{2}\bar{A}^{mnop}\epsilon_{mn}\epsilon_{op} \;, \quad \bar{A}^{mnop}=\frac{E}{1+\nu}\left(\frac{\nu}{1-\nu}\bar{a}^{mn}\bar{a}^{op}+\frac{1}{2}\bar{a}^{mo}\bar{a}^{np}+\frac{1}{2}\bar{a}^{mp}\bar{a}^{no}\right)
\end{equation}
where $\bar{A}^{mnop}$ is the elasticity tensor for an isotropic material \cite{landau:1986a} and $\bar{a}^{ij} = \bar{a}_{ij}^{-1}$. The stretching energy density can be simplified to 
    \begin{align}\label{eq:ws}
    w_s=\frac{Eh}{2(1+\nu)}&\left(\frac{1}{1-\nu}((1+\bar{e}_c)^4\epsilon_{11}^2+(1+\bar{e}_s)^4\epsilon_{22}^2)     +(1+\bar{e}_c)^2(1+\bar{e}_s)^2\left(\frac{2\nu}{1-\nu}\epsilon_{11}\epsilon_{22} + 2\epsilon_{12}^2\right)\right).
    \end{align}

We take the variation of the stretching energy
\begin{equation}\label{eq:Ws}
W_s = \iint w_s \d\alpha_1 \d\alpha_2
\end{equation}
with respect to the position $\mathbf{r}$ to obtain the stretching force per unit material area, i.e. $-\delta w_s/\delta \mathbf{r}$.

Integrating by parts to move derivatives off of~$\delta\mathbf{r}$ terms, we obtain 
\begin{align} 
\delta W_s=&\frac{Eh}{1+\nu}\oint\left[ \left(\frac{1}{1-\nu}(1+\bar{e}_c)^4 \epsilon_{11}\left(\frac{\partial\mathbf{r}}{\partial \alpha_1}\cdot\delta \mathbf{r}\right)+\frac{\nu}{1-\nu}(1+\bar{e}_c)^2(1+\bar{e}_s)^2 \epsilon_{22}\left(\frac{\partial\mathbf{r}}{\partial \alpha_1}\cdot\delta \mathbf{r}\right) \right.\right.\nonumber\\
&\hspace{.7cm}\left.+(1+\bar{e}_c)^2(1+\bar{e}_s)^2\epsilon_{12}\left(\frac{\partial\mathbf{r}}{\partial \alpha_2}\cdot\delta \mathbf{r}\right) \right)\upsilon_1\nonumber\\
&\hspace{.7cm}+\left. \left( \frac{1}{1-\nu}(1+\bar{e}_s)^4 \epsilon_{22}\left(\frac{\partial\mathbf{r}}{\partial \alpha_2}\cdot\delta \mathbf{r}\right)+\frac{\nu}{1-\nu}(1+\bar{e}_c)^2(1+\bar{e}_s)^2 \epsilon_{11}\left(\frac{\partial\mathbf{r}}{\partial \alpha_2}\cdot\delta \mathbf{r}\right) \right.\right.\nonumber\\
&\hspace{.7cm}\left.\left.+(1+\bar{e}_c)^2(1+\bar{e}_s)^2\epsilon_{12}\left(\frac{\partial\mathbf{r}}{\partial \alpha_1}\cdot\delta \mathbf{r}\right) \right)\upsilon_2 \right]\,\d\sigma \nonumber\\
&-\frac{Eh}{1+\nu} \iint \left[\frac{\partial}{\partial \alpha_1}  \left(\frac{1}{1-\nu}(1+\bar{e}_c)^4 \epsilon_{11}\frac{\partial\mathbf{r}}{\partial \alpha_1}+\frac{\nu}{1-\nu}(1+\bar{e}_c)^2 (1+\bar{e}_s)^2\epsilon_{22}\frac{\partial\mathbf{r}}{\partial \alpha_1} \right.\right.\nonumber\\
&\hspace{.7cm}\left.\left.+(1+\bar{e}_c)^2(1+\bar{e}_s)^2\epsilon_{12}\frac{\partial\mathbf{r}}{\partial \alpha_2} \right)\right.    \nonumber\\
&\hspace{.7cm} \left. +\frac{\partial}{\partial \alpha_2} \left( \frac{1}{1-\nu}(1+\bar{e}_s)^4 \epsilon_{22}\frac{\partial\mathbf{r}}{\partial \alpha_2}+\frac{\nu}{1-\nu}(1+\bar{e}_c)^2(1+\bar{e}_s)^2 \epsilon_{11}\frac{\partial\mathbf{r}}{\partial \alpha_2}\right.\right.\nonumber\\
&\hspace{.7cm}\left.\left.+(1+\bar{e}_c)^2(1+\bar{e}_s)^2\epsilon_{12}\frac{\partial\mathbf{r}}{\partial \alpha_1}\right)  \right]\cdot\delta\mathbf{r}\,\d \alpha_1\d \alpha_2.\label{dWs}
\end{align}
The first integral in (\ref{dWs}) can be used to obtain the free edge boundary conditions. It is a boundary integral with respect to $\d\sigma$, arc length along the boundary in the $\alpha_1$-$\alpha_2$ plane;  $(\upsilon_1,\upsilon_2)$ is the outward normal in this plane. The integrand in the second integral in (\ref{dWs})
gives the stretching force per unit material area,
\begin{align}
\mathbf{f}_s=-\frac{\delta w_s}{\delta\mathbf{r}}=&\frac{Eh}{1+\nu} \left[\frac{\partial}{\partial \alpha_1}  \left(\frac{1}{1-\nu}(1+\bar{e}_c)^4 \epsilon_{11}\frac{\partial\mathbf{r}}{\partial \alpha_1}+\frac{\nu}{1-\nu}(1+\bar{e}_c)^2 (1+\bar{e}_s)^2\epsilon_{22}\frac{\partial\mathbf{r}}{\partial \alpha_1} \right.\right.\nonumber\\
&\hspace{.7cm}\left.\left.+(1+\bar{e}_c)^2(1+\bar{e}_s)^2\epsilon_{12}\frac{\partial\mathbf{r}}{\partial \alpha_2} \right)\right.    \nonumber\\
&\hspace{.7cm} \left. +\frac{\partial}{\partial \alpha_2} \left( \frac{1}{1-\nu}(1+\bar{e}_s)^4 \epsilon_{22}\frac{\partial\mathbf{r}}{\partial \alpha_2}+\frac{\nu}{1-\nu}(1+\bar{e}_c)^2(1+\bar{e}_s)^2 \epsilon_{11}\frac{\partial\mathbf{r}}{\partial \alpha_2}\right.\right.\nonumber\\
&\hspace{.7cm}\left.\left.+(1+\bar{e}_c)^2(1+\bar{e}_s)^2\epsilon_{12}\frac{\partial\mathbf{r}}{\partial \alpha_1}\right)  \right].\label{eq:3dtensionApp}
\end{align}

\section{Stability boundaries}\label{app:StabilityBoundaries}

\begin{figure}[H]
    \centering
    \includegraphics[width=\textwidth]{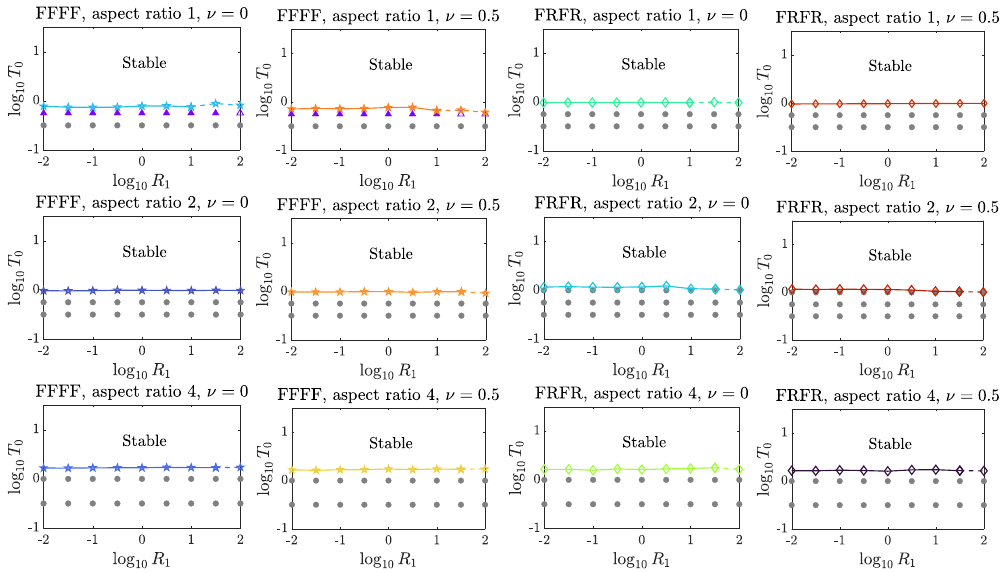}
    \caption{Classification of instability types for the FFFF and FRFR boundary conditions with Poisson ratio 0 and 0.5, and aspect ratio one, two, and four. The colored lines with the star and diamond markers, respectively, separate the regions where the FFFF and FRFR membranes are stable and unstable. The purple triangles mark where membranes become unstable through flutter and divergence and the gray dots mark where membranes lose stability by divergence. Empty triangle markers correspond to cases where the membranes lose stability with oscillations but without a well-defined frequency of oscillation. For the stability results we use the spanwise symmetric initial condition with $\sigma=10^{-5}$.}
    \label{fig:stability}
\end{figure}

\begin{figure}[H]
    \centering
    \includegraphics[width=\textwidth]{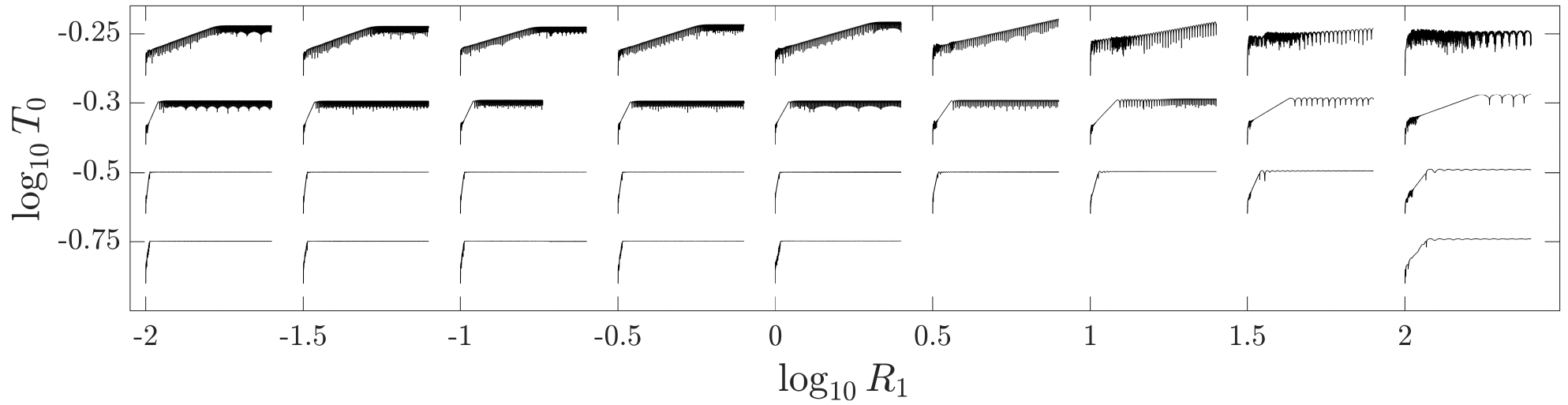}
    \caption{$\log_{10}|z_{\mathrm{center}}(t)|$ in $R_1$-$T_0$ space for FFFF membranes with aspect ratio one and $\nu=0$. The 
    horizontal $t$ and vertical $\log_{10}|z_{\mathrm{center}}|$ axes are omitted. The range of $t$ is $[0, 1000]$ in most cases and the vertical scale is normalized by the maximum of $\log_{10}|z_{\mathrm{center}}(t)|$ over the time range shown.}\label{fig:zcenterFlutterAndDiv}
\end{figure}

Here figure~\ref{fig:stability} marks $(R_1,T_0)$ pairs where the small-amplitude instability corresponds to divergence or flutter with divergence, for each of the two boundary conditions, Poisson ratios 0 and 0.5, and
membrane aspect ratios one, two, and four. Divergence (without flutter) is the most common instability type, but flutter with divergence is also observed close to the stability boundary for FFFF membranes with aspect ratio one.
In figure~\ref{fig:zcenterFlutterAndDiv} we show in $R_1$-$T_0$ space the $z_{\mathrm{center}}(t)$ plots on a log scale for FFFF membranes with aspect ratio one and $\nu=0$. At the largest $T_0$, $10^{-0.25}$ (close to the stability boundary), the membranes become unstable by divergence with flutter. This corresponds to exponential growth with a sinusoidal oscillatory component. The sign changes during the oscillations are shown by the regularly spaced vertical asymptotes in the graphs of the logarithm of $z_{\mathrm{center}}$ during the exponential growth, where the envelopes of the graphs have an upward slope. When the deflections reach large amplitude, the periodic oscillation remains, but the frequency increases by about 60--100\% depending on $R_1$. 
At the next smaller $T_0$, $10^{-0.3}$,  the duration of exponential growth is much shorter and the sign changes are not seen, corresponding to divergence without flutter (i.e., without oscillations). However, sinusoidal oscillation occur at large amplitude, but with a slightly smaller frequency than at $T_0 = 10^{-0.25}$.
At $T_0 = 10^{-0.5}$, divergence without oscillation is seen and no oscillations are seen at large amplitude.
At $T_0 = 10^{-0.75}$, irregular (nonsinusoidal) oscillations are seen during the exponential growth regime; the type of instability is unclear here. 

\bibliographystyle{unsrt}
\bibliography{biblio.bib}

\end{document}